\documentclass[final,leqno,onefignum,onetabnum]{siamltex1213}
\usepackage{amsmath, graphicx,amssymb,epsfig,psfrag,subfigure,color}

\newcommand{\beq}{\begin{equation}}
\newcommand{\eeq}{\end{equation}}
\newcommand{\bpm}{\begin{pmatrix}}
\newcommand{\epm}{\end{pmatrix}}
\newcommand{\beqa}{\begin{eqnarray}}
\newcommand{\eeqa}{\end{eqnarray}}
\newcommand{\beqas}{\begin{eqnarray*}}
\newcommand{\eeqas}{\end{eqnarray*}}

\renewcommand{\d}{\mathrm{d}}

\newcommand{\CO}{\mathcal{O}}

\newcommand{\CF}{\mathcal{F}}

\newcommand{\sech}{\text{sech}}

\newcommand{\pdhfrac}[2]{\mathchoice{\frac{#1}{#2}}{#1/#2}{#1/#2}{#1/#2}}

\newcommand{\pd}[2]{\pdhfrac{{\partial}#1}{{\partial}#2}}

\newcommand{\vb}{\textbf{b}}

\def\Xint#1{\mathchoice
{\XXint\displaystyle\textstyle{#1}}%
{\XXint\textstyle\scriptstyle{#1}}%
{\XXint\scriptstyle\scriptscriptstyle{#1}}%
{\XXint\scriptscriptstyle\scriptscriptstyle{#1}}%
\!\int}
\def\XXint#1#2#3{{\setbox0=\hbox{$#1{#2#3}{\int}$ }
\vcenter{\hbox{$#2#3$ }}\kern-.6\wd0}}

\def\dashint{\Xint-}

\title{HOMOGENISATION OF A ROW OF DISLOCATION DIPOLES FROM DISCRETE DISLOCATION DYNAMICS\thanks{This work was partly supported by EPSRC through grant EP/D048400/1, and by the Hong Kong Research Grants Council through General Research Fund 606313}.}

\author{Stephen Jonathan Chapman\thanks{Mathematical Institute, University of Oxford, Andrew Wiles Building, Radcliffe Observatory Quarter, Woodstock Road, Oxford, OX2 6GG, UK.
(\email{chapman@maths.ox.ac.uk}).} \and Yang Xiang$^{\text{\ddag}}$ \and Yichao Zhu\thanks{Department of Mathematics, the Hong Kong University of Science and Technology, Clear Water Bay, Kowloon, Hong Kong, China.
(\email{maxiang@ust.hk} and \email{mayczhu@ust.hk}).}}

\begin{document}
\maketitle
\slugger{siap}{xxxx}{xx}{x}{x--x}

\begin{abstract}
  Conventional discrete-to-continuum approaches have seen their limitation in describing the collective behaviour of the multi-polar configurations of dislocations, which are widely observed in crystalline materials. The reason is that dislocation dipoles, which play an important role in determining the mechanical properties of crystals, often get smeared out when traditional homogenisation methods are applied. To address such difficulties, the collective behaviour of a row of dislocation dipoles is studied by using matched asymptotic techniques. The discrete-to-continuum transition is facilitated by introducing two field variables respectively describing the dislocation pair density potential and the dislocation pair width. It is found that the dislocation pair width evolves much faster than the pair density. Such hierarchy in evolution time scales enables us to describe the dislocation dynamics at the coarse-grained level by an evolution equation for the slowly varying variable (the pair density) coupled with an equilibrium equation for the fast varying variable (the pair width). The time-scale separation method adopted here paves a way for properly incorporating dipole-like (zero net Burgers vector but non-vanishing) dislocation structures, known as the statistically stored dislocations (SSDs) into macroscopic models of crystal plasticity in three dimensions. Moreover, the natural transition between different equilibrium patterns found here may also shed light on understanding the emergence of the persistent slip bands (PSBs) in fatigue metals induced by cyclic loads.
\end{abstract}



\pagestyle{myheadings}
\thispagestyle{plain}
\markboth{HOMOGENISATION OF A ROW OF DISLOCATION DIPOLES}{S. J. CHAPMAN, Y. XIANG, AND Y. C. ZHU}

\section{Introduction}
\begin{figure}[!ht]
  \centering
  \subfigure[Channel-vein structure]{ \includegraphics[width=.35\textwidth]{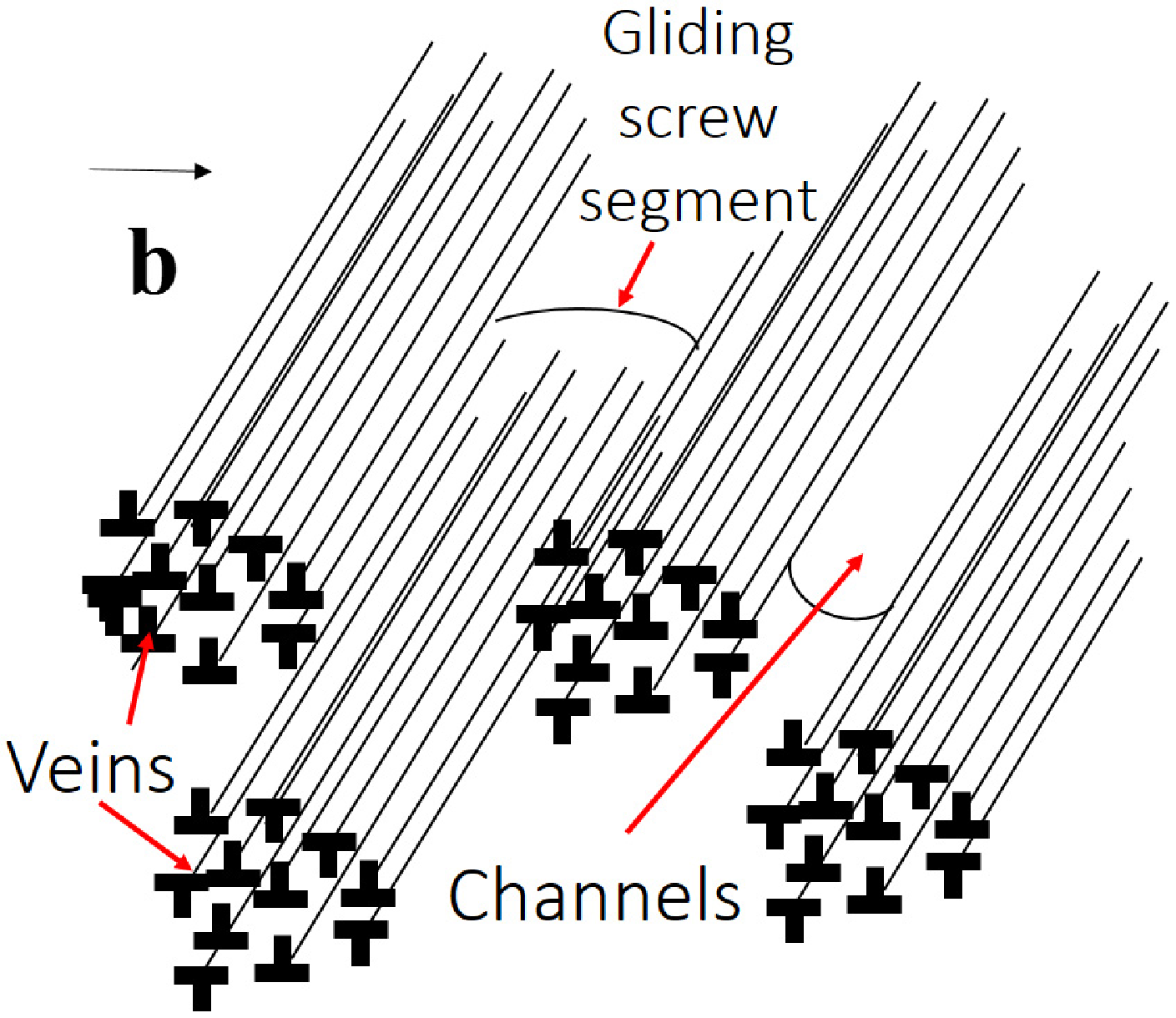}}
  \subfigure[PSB structure]{ \includegraphics[width=.35\textwidth]{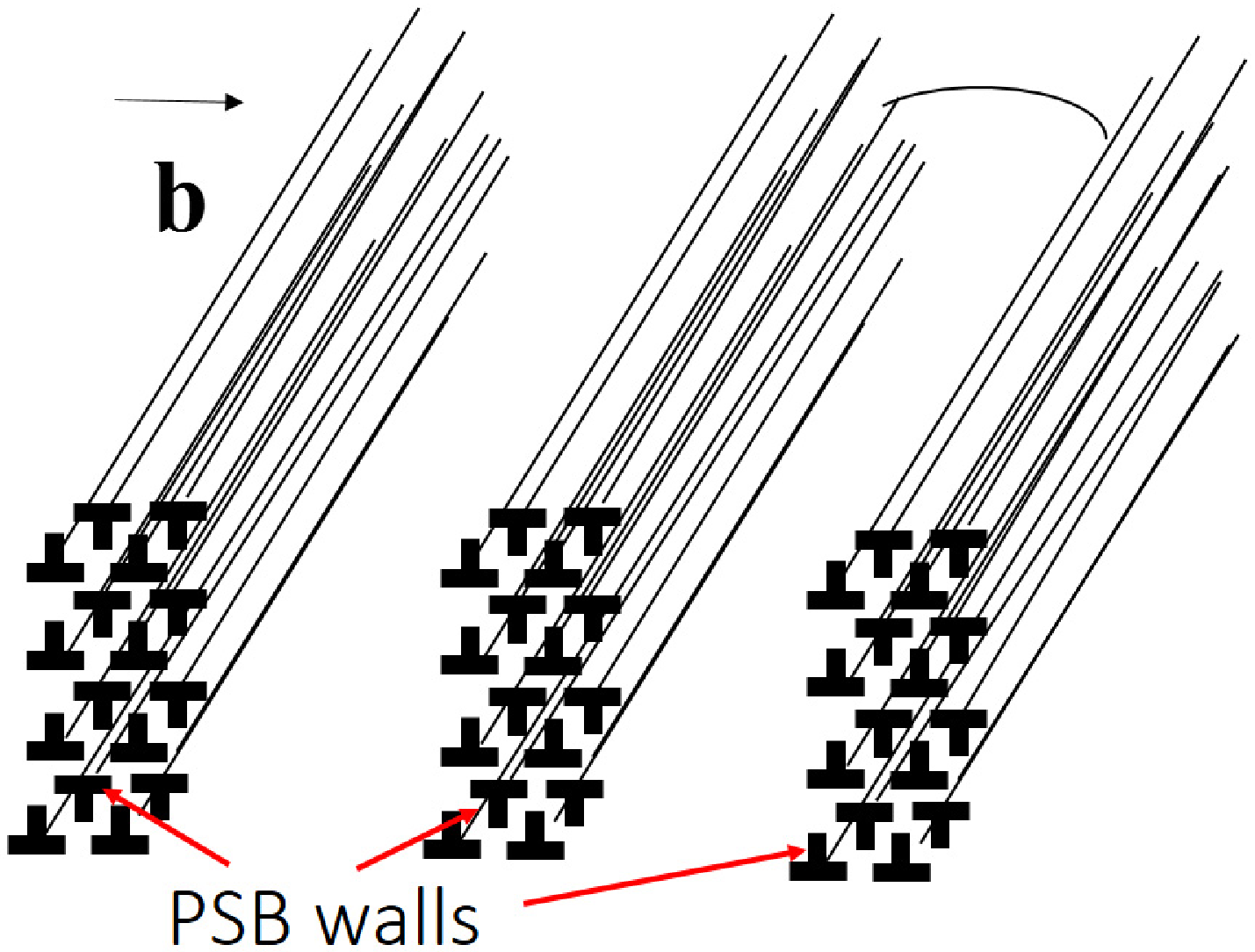}}
  \caption{Dislocation patterns in the early stage of metal fatigue induced by cyclic loads. \label{fig_vein_PSBs}}
\end{figure}

It is well known that the plastic deformation of crystalline materials is carried by a large number of atomistic line defects, i.e. dislocations. Hence macroscopic models of crystal plasticity can be established by formulating the dynamics of many dislocations. As an idealised (but also practically useful) case, the dynamics of straight and mutually-parallel dislocations have been intensively studied. These translationally invariant dislocations can be treated as ``poles'' on one of the planes perpendicular to all dislocation lines. These straight dislocations, like electrical charges, have signs depending on their line directions with respect to the slip direction, known as the Burgers vector. Abundant experimental evidence suggests that a good understanding of the collective behaviour of many straight dislocations is important for controlling the mechanical properties of crystals. One example is found inside a single-crystalline fatigued copper specimen induced by cyclic loads \cite{Mughrabi_1979}. Before the saturation point is reached, the inner configuration of the copper specimen takes a ``channel-vein'' structure as shown in Fig.~\ref{fig_vein_PSBs}(a).
A vein consists of many almost straight and closely spaced edge dislocations and the veins are separated by channels where the dislocation density is relatively low. Beyond the saturation point, a characteristic ladder-shape structure (known as persistent slip bands (PSBs)) forms as shown in Fig.~\ref{fig_vein_PSBs}(b). The walls of the ladders also consist of straight edge dislocations. The mechanism governing the transition from the channel-vein to PSB structures is still unclear, and a study of the collective behaviour of edge dipoles can be of great help to its understanding.

One way to reveal the role played by these straight dislocations during the formation of PSBs, is by using two-dimensional (2D) discrete dislocation dynamical (DDD) models, where all dislocations are tracked individually (e.g. \cite{Brinckmann2004}). Nevertheless, it is still difficult to get a clear idea of the mechanism that governs dislocation pattern formation in crystals from DDD simulations. Hence there is still a need to investigate the dynamics of dislocations at the continuum level, where materials substructures are described by a dislocation density distribution. In principle, a dislocation-density-based continuum model should be obtained through a rigorous averaging of its underlying 2D DDD model. However, existing discrete-to-continuum approaches struggle to upscale multi-polar configurations of straight dislocations. The reason is as follows. At room temperature, dislocations (of edge type) are in general constrained in their own slip planes. As a result, a positively defined edge dislocation and a nearby negatively defined edge dislocation which is not on the same slip plane tend to lock each other by forming a pair of dipole rather than to annihilate each other. Since the locking stress between the two components of a dislocation dipole scales with the intra-dipolar spacing $r$ by $1/r$, a relatively large externally applied stress is needed in order to mobilise the constituent dislocations of a dipole. Hence the presence of dislocation dipoles may effectively increase the strength of a crystal. When traditional homogenisation methods are applied, however, dipoles, despite being crucial in determining the material mechanical properties, average to zero and they play no role in the continuum approximation. Owing to this, traditional homogenisation techniques are only applicable when investigating the collective behaviour of many dislocations of the same sign, i.e. the geometrically necessary dislocations (GNDs) (e.g. \cite{Geers2013, Ock1983, Voskoboinikov2007JMPS}). The collective behaviour of an arbitrary multipolar configuration of dislocations is only considered in a phenomenological or statistical manner \cite{Schulz2014, Groma1997, Groma2003}. There have also been works where each dipole pair is treated as one object so that the traditional homogenisation method can be applied \cite{Hall2010}. As shown by our analysis, this only works for the case where the slip plane spacing is much smaller than the inter-dipolar spacing.

Capturing dipole-like structures at the continuum level is also a bottom neck problem for establishing a three-dimensional (3D) dislocation-density-based continuum theory of plasticity. The density distribution of GNDs in 3D space, where dislocations can be curved, is represented by the Nye's dislocation density tensor \cite{Nye1953}, which only accommodates the gradient of (macroscopic) plastic strains. One missing part is the role played by statistically stored dislocations (SSDs), whose physical dimensions are too small to be distinguished from a dislocation-free state in the Nye's dislocation density tensor. Similar as dislocation dipoles discussed above, some SSD structures also play a role in determining the (macroscopic) plastic properties of crystals. During the past two decades, many valuable works have been done in order to improve the framework based on the Nye's dislocation density tensor (e.g. \cite{Acharya2001, Arsenlis2002, ElAzab2000, Evers_JMPS2004, Tighe1993, Hochrainer2007, Sandfeld2011}), but the formulation of SSDs at the continuum level is either phenomenological or statistical up to date. There are other dislocation configurations that are not properly included in the framework based on the Nye's dislocation density tensor, such as dislocation interactions with other types of crystalline defects (e.g. Frank-Read sources, grain boundaries). Therefore, a pivotal question to be answered for establishing a solid dislocation-density-based theory of plasticity is, ``how should SSDs as well as other structures missing in the framework for GNDs be properly formulated on a coarse-grained scale?'' Part of this question has been answered through the establishment of a continuum model of plasticity, where a set of dislocation density potential functions (DDPFs) are employed to represent the dislocation substructures on a single slip plane \cite{Xiang2009_JMPS, ZhuXH2010, Zhu2014_IJP} and in three-dimensional space \cite{Zhu_continuum3D}. The micro-scale mechanisms that are well incorporated into the continuum model characterised by DDPFs are the dislocation line tension effect \cite{Xiang2009_JMPS}, the grain boundary structures \cite{Xiang_GB2014} and the operation of dislocation sources of the Frank-Read type \cite{Zhu2014_IJP}. The hints of how to rigorously incorporate SSDs at the continuum level can be found from the analysis presented in the current paper.

Motivated by these issues, the collective behaviour of a row of dislocation dipoles is studied here. The discrete-to-continuum transition is facilitated by the introduction of two field variables respectively describing the dislocation pair density potential and the dislocation pair width. By using asymptotic analysis, we derive coupled evolution equations for these two field variables. Actually we show that the time scales associated with the evolution of the two field variables are different. The dislocation pair width, which moves in response to the resolved shear stress at leading order, varies on a time scale much shorter than that associated with the dislocation pair density, which moves in response to the ``stress gradient'' (coming from the resolved shear stress at the next order). Hence if viewed on the slower scale, fast-varying mechanisms take place so quickly that only their steady (or equilibrium) states need to be taken into account. As a result, the collective behaviour of a row of dislocation dipoles at the continuum level can be described by an evolution equation for the slowly varying dislocation pair density coupled with an equilibrium equation for the dislocation pair width. Such discrete-to-continuum approaches asymptotically separating active processes according to their associated time scales may pave a way for the incorporation of SSDs at the continuum level. Moreover, a transition between two distinct dipolar patterns due to instability, which was originally discovered in periodically distributed dipoles \cite{Zhu_2Ddipoles2014}, is also seen here, and the transition may have some role to play in understanding the formation of PSBs.

The paper is arranged as follows. The governing equations for the DDD model, which we take as our reference model, are written down in \S\ref{Sec_DDD}. After the introduction of the variables needed for the discrete-to-continuum transition in \S\ref{Sec_prepare}, we derive for the asymptotic expressions of the resolved stress field in \S\ref{Sec_stress}. Then the governing equations for equilibrium states and the dynamics at the continuum level are presented in \S\ref{Sec_continuum_model}. In \S\ref{Sec_equilibria}, the equilibrium states at the continuum level are further analysed and a natural transition between different equilibrium patterns is found. In \S\ref{Sec_continuum_DDD_comp}, the accuracy and efficiency of the derived continuum model are studied. The article concludes with further discussion in \S\ref{Sec_conclusion}.

\section{Dynamics at the level of discrete dislocations\label{Sec_DDD}}
\begin{figure}[!ht]
\centering
\includegraphics[width=.65\textwidth]{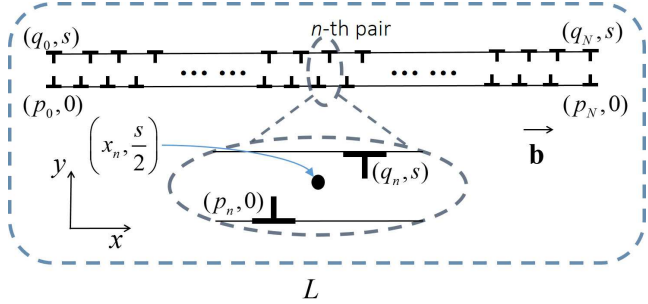}
\caption{The $x$-$y$ plane is one of the planes perpendicular to all dislocation lines ,and $\vb$ is the Burgers vector. The numbers of positive and negative dislocations are identically $N+1$. All positively oriented dislocations are located on one slip plane, which degenerates to the $x$-axis here, and all negatively oriented dislocations are put on another slip plane at a distance of $s$ from the $x$-axis (given by $y=s$). The $n$-th dislocation pair consists of the $n$-th positive and negative dislocations, whose locations are set to be at $(p_n,0)$ and $(q_n,s)$, respectively. The $x$-coordinate for the center of the $n$-th pair is denoted by $x_n$ given by Eq.~\eqref{xn_def}. Here the length of the domain of interest $L$ equals $1$ after non-dimensionalisation. \label{fig_problem_set_up}}
\end{figure}
Here we consider the case of a single slip system associated with the Burgers vector denoted by $\vb$, and all dislocations here are straight, mutually parallel and of edge type. The problem is thus reduced to one of the planes that are orthogonal to all dislocation tangents. Here the plane of interest is set to be the $x$-$y$ plane as shown in Fig.~\ref{fig_problem_set_up}. If we choose $\vb=(b,0)$ with $b>0$, each dislocation can thus be treated as a signed point in $x$-$y$ plane. Here we set a dislocation with its line direction pointing outward the paper plane (see \cite{Hirth1982} for details) to be a ``positive dislocation'' and denoted by ``$\bot$''. A dislocation with its line direction pointing inward the paper plane is set to be a ``negative dislocation'' and denoted by ``$\top$''.

The configuration we consider is shown in Fig.~\ref{fig_problem_set_up}. There are $N+1$ positive dislocations lying on the slip plane characterised by the $x$-axis, while $N+1$ negative dislocations are put on another slip plane at a distance of $s$ from the $x$-axis. The $n$-th dislocation pair is set to consist of the $n$-th positive and negative dislocations, which are located at $(p_n,0)$ and $(q_n,s)$, respectively.

Concerning dislocation motion, we employ a dislocation mobility law, which only allows dislocations (of edge types) to glide within their slip plane at a speed proportional to their on-site resolved shear stress. Under this rule, the motion of the $n$-th positive dislocation is governed by
\beq \label{mobility_law_pos}
v_n^+ = \frac{\d p_n}{\d t} = m_{\text{g}} b (\tau_{\text{int}}(p_n,0) + \tau_{\text{ext}}(p_n,0)),
\eeq
where $v_n^+$ denotes the speed of the $n$-th positive dislocation along $x$-direction; $\tau_{\text{int}}(x,y)$ is the internal resolved shear stress field at $(x,y)$ resulting from the dislocation-dislocation interactions; $\tau_{\text{ext}}(x,y)$ denotes the externally applied resolved shear stress at $(x,y)$; $m_{\text{g}}$ is the dislocation glide coefficient; $b=|\vb|$.

To facilitate further analysis, we consider the problem in a non-dimensional sense, that is, all spatial variables are scaled with $L$; all stress components are scaled with $\mu Nb/(2\pi(1-\nu)L)$ and time $t$ is scaled with $2\pi (1-\nu)L^2/(\mu m_{\text{g}}Nb^2)$, where $\mu$ and $\nu$ are the shear modulus and Poisson's ratio, respectively. Hence the non-dimensional version of Eq.~\eqref{mobility_law_pos} becomes
\beq \label{mobility_law_pos_dimensionless}
\frac{\d p_n}{\d t} = \tau_{\text{int}}(p_n,0) + \tau_{\text{ext}}(p_n,0) = \tau_{\text{tot}}(p_n,0),
\eeq
where $\tau_{\text{tot}}$ denotes the (non-dimensional) total resolved shear stress field. Similarly, the (non-dimensional) gliding speed of the $n$-th negative dislocation is governed by
\beq \label{mobility_law_neg_dimensionless}
\frac{\d q_n}{\d t} = -\tau_{\text{int}}(q_n,s) - \tau_{\text{ext}}(q_n,s) = -\tau_{\text{tot}}(q_n,s).
\eeq
A comparison between Eqs.~\eqref{mobility_law_pos_dimensionless} and \eqref{mobility_law_neg_dimensionless} suggests that a positive and a negative dislocation move in opposite directions under the same resolved shear stress field.

The (non-dimensional) internal resolved shear stress field $\tau_{\text{int}}$ is calculated by the superposition of the resolved shear stresses due to all individual dislocations \cite{Hirth1982}:
\beq \label{tau_internal_pos_n_dimensionless}
\tau_{\text{int}}(p_n,0) = \frac1{N}\sum_{\substack{j=0 \\ j\neq n}}^N \frac1{p_n-p_j} - \frac1{N}\sum_{j=0}^N \frac{(p_n-q_j)((p_n-q_j)^2 - s^2)}{((p_n-q_j)^2 + s^2)^2}
\eeq
and
\beq \label{tau_internal_neg_n_dimensionless}
\tau_{\text{int}}(q_n,s) = \frac1{N}\sum_{j=0}^N \frac{(q_n-p_j)((q_n-p_j)^2 - s^2)}{((q_n-p_j)^2 + s^2)^2} - \frac1{N}\sum_{\substack{j=0 \\ j\neq n}}^N \frac1{q_n-q_j}.
\eeq
The dynamics at the level of discrete dislocations is thus given by Eqs.~\eqref{mobility_law_pos_dimensionless} - \eqref{tau_internal_neg_n_dimensionless}, which form a closed system of ordinary differential equations for the $2(N+1)$ unknowns $\{p_n\}_{n=0}^N$ and $\{q_n\}_{n=0}^N$.

\section{Preparation for discrete-to-continuum transition\label{Sec_prepare}}
Usually the number of dislocations in crystals is very large. Hence it is sensible to consider the collective behaviour of the system governed by Eqs.~\eqref{mobility_law_pos_dimensionless} - \eqref{tau_internal_neg_n_dimensionless}. Mathematically, this can be achieved by examining the asymptotic behaviour of the system as $N\rightarrow\infty$. The expected outcomes are the evolution equations of some continuously defined variables that characterise the dislocation substructures. In this section, we will introduce the field variables needed for the discrete-to-continuum transition.

Given a large $N$, the length scale associated with the discrete dislocation dynamical model given by Eqs.~\eqref{mobility_law_pos_dimensionless} to \eqref{tau_internal_neg_n_dimensionless} is the spacing of neighbouring discrete dislocations, i.e. $\CO(1/N)$, so that individual dislocations can be observed. We now want to describe the same dynamical relation by a model associated with an $\CO(1)$ length scale, where a continuous dislocation density distribution is considered rather than isolated dislocations.

To facilitate such a transition, we first define $x_n$ to be the $x$ coordinate for the center of the $n$-th dislocation pair as shown in Fig~\ref{fig_problem_set_up}
\beq \label{xn_def}
x_n = \frac{p_n+q_n}{2}.
\eeq
Then we introduce a continuous function of (non-dimensional) time and space denoted by $\zeta(t,x)$, such that
\beq \label{zeta_def}
\frac{\zeta(t,x_n)}{N} = q_n-p_n.
\eeq
Here $\zeta(t,x)$ is a field variable defined for $x\in[0,1]$, and its value at $x_n$ measures the width of $n$-th dislocation pair scaled by $N$. Since the spacings of neighbouring dislocations are $\CO(1/N)$, $\zeta(t,x)\sim\CO(1)$ as $N\rightarrow\infty$. At the continuum level, $\zeta$ is employed to characterise the local pattern of dislocation dipoles.

Throughout the paper, a subscript $n$ or $j$ affiliated with a field variable such as $\zeta$ indicates that the field is evaluated at $x=x_n$ or at $x=x_j$, respectively; for example, $\zeta_n = \zeta(t,x_n)$. Here we consider the case where $\zeta\in[0,1/2]$, and the properties for $\zeta\in(-1/2,0)$ can be studied likewise.

From Eqs.~\eqref{xn_def} and \eqref{zeta_def}, $p_n$ and $q_n$ can be expressed in terms of $x_n$ and $\zeta_n$ respectively by
\beq \label{pqn_from_xnzetan}
p_n = x_n - \frac{\zeta_n}{2N},\qquad
q_n = x_n + \frac{\zeta_n}{2N}.
\eeq

We now introduce another field variable, the dislocation pair density potential $\phi(t,x)$, such that
\beq \label{DDPF_def}
\phi_n = \phi(t,x_n) = \frac{n}{N};
\eeq
this definition is by direct analogy with the dislocation density potential functions defined in \cite{Xiang2009_JMPS} or \cite{Zhu_continuum3D}. It can be shown by following the same argument presented in \cite{Xiang2009_JMPS} that the density distribution of the dislocation pairs denoted by $\rho$ can be calculated by $\rho = \partial \phi/\partial x$. Throughout this paper, the inputs $(t,x)$ for $\phi$ and $\zeta$ are omitted if no ambiguities are caused. Moveover, a dash is added to a variable to denote its derivative with respect to $x$, for example $\rho=\phi'$.

Here we only consider the case when $s$, the (non-dimensional) spacing between the two slip planes, is $\CO(1/N)$, as $N\rightarrow\infty$. This implies that $s$ can be rescaled by
\beq \label{s_scale}
s = \frac{S}{N},
\eeq
where $S\sim\CO(1)$. When $s\sim \CO(1)$, the interaction between dislocations from different slip planes become long-range, and the configurations can be studied by applying conventional homogenisation approaches.

At the continuum level, the dislocation substructures are expected to be described by the two field variables $\phi$ and $\zeta$ and the goal now is to look for their governing equations by taking the asymptotic limit $N\rightarrow\infty$ of Eqs.~\eqref{mobility_law_pos_dimensionless} - \eqref{tau_internal_neg_n_dimensionless}.

\section{Asymptotic behaviour of the resolved shear stress field\label{Sec_stress}}
To accomplish the discrete-to-continuum transition we use the following procedure. Given a quantity defined in a discrete sense, we first asymptotically express the values at $(p_n,0)$ and $(q_n,s)$ by functions of $x_n$, for any integer $n\in[0,N]$. In this way the equations at the discrete level can be transformed into asymptotic equations for $\phi$ and $\zeta$, which only hold at every $x_n$. Then by using the fact that $x_n$ is densely distributed throughout the whole domain, we replace $x_n$ by $x$ to turn the obtained equations to corresponding integro-differential equations of $\phi$ and $\zeta$, which hold for all $x$.

\begin{figure}[!ht]
  \centering
  \includegraphics[width=.7\textwidth]{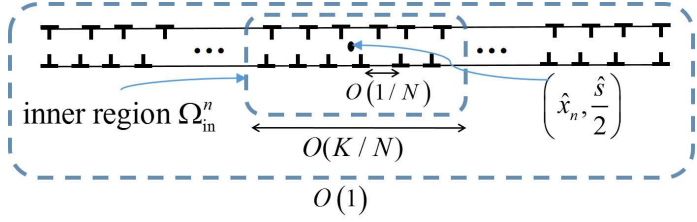}
  \caption{Given the $n$-th dislocation pair, the $x$-coordinates of the centers of its $2K$ neighbouring pairs fall inside an interval, defined to be the inner region $\Omega_{\text{in}}^n$, whose size is $\CO(K/N)$. Mathematically, this interval is given by Eq.~\eqref{inner_region_def}. The outer region is defined to be the interval into which the $x$-coordinates of the centers of all other dislocation pairs fall. Mathematically, it is given by Eq.~\eqref{outer_region_def}. \label{fig_region_split}}
\end{figure}

Following this strategy, we start by considering the asymptotic behaviour of the internal resolved shear stress $\tau_{\text{int}}(p_n,0)$ and $\tau_{\text{int}}(q_n,s)$, given by Eq.~\eqref{tau_internal_pos_n_dimensionless} and \eqref{tau_internal_neg_n_dimensionless}, respectively as $N\rightarrow\infty$. First, an interval $\Omega_{\text{in}}^n$ is introduced to the $n$-th dislocation pair, such that the $x$-coordinates of the centers of its $2K$ neighbouring pairs fall inside $\Omega_{\text{in}}^n$ as shown in Fig.~\ref{fig_region_split}.
The number $K$ here satisfies $1\ll K \ll N$.
Throughout this paper, $\Omega_{\text{in}}^n$ defined in this way is termed as the ``inner region''. Mathematically, $\Omega_{\text{in}}^n$ is represented by
\beq \label{inner_region_def}
\Omega_{\text{in}}^n = \left\{x\left| \left|\phi(t, x)-\frac{n}{N}\right|\le \frac{K}{N} \right.\right\}.
\eeq
It can be seen that the length of $\Omega_{\text{in}}^n$ is $\CO(K/N)$.
Similarly we define the ``outer region'' by
\beq \label{outer_region_def}
\Omega_{\text{out}}^n = \left\{x\left| \left|\phi(t, x)-\frac{n}{N}\right| > \frac{K}{N} \right.\right\},
\eeq
which contains the $x$-coordinate of the centers of all other dislocation pairs. Then we estimate $\tau_{\text{int}}(p_n,0)$ in  Eq.~\eqref{tau_internal_pos_n_dimensionless} by decomposing it into two parts
\beq \label{tau_p_split}
\tau_{\text{int}}(p_n,0) = \tau_{\text{int}}^{\text{in}}(p_n,0) + \tau_{\text{int}}^{\text{out}}(p_n,0),
\eeq
where $\tau_{\text{int}}^{\text{in}}(p_n,0)$ denotes the resolved shear stress due to all dislocations associated with the inner region $\Omega_{\text{in}}^n$:
\beq \label{tau_p_inner}
\tau_{\text{int}}^{\text{in}}(p_n,0) = \frac1{N}\sum_{\substack{j=n-K \\ j\neq n}}^{n+K} \frac1{p_n-p_j} - \frac1{N}\sum_{j=n-K}^{n+K} \frac{(p_n-q_j)((p_n-q_j)^2 - s^2)}{((p_n-q_j)^2 + s^2)^2}
\eeq
and $\tau_{\text{int}}^{\text{out}}(p_n,0)$ denotes the resolved shear stress due to all dislocations associated with $\Omega_{\text{out}}^n$:
\beq \label{tau_p_outer}
\tau_{\text{int}}^{\text{out}}(p_n,0) = \frac1{N}\sum_{\substack{0\le j<n-K\\n+K < j\le N}} \left(\frac1{p_n-p_j} - \frac{(p_n-q_j)((p_n-q_j)^2 - s^2)}{((p_n-q_j)^2 + s^2)^2}\right).
\eeq
It is worth noting that the decomposition suggested by Eq.~\eqref{tau_p_split} only holds for dislocation pairs that are not too close to the boundaries, i.e. $K< n < N-K$.

We will perform an inner and an outer region approximation to calculate the asymptotic limit of $\tau_{\text{int}}^{\text{in}}(p_n,0)$ and $\tau_{\text{int}}^{\text{out}}(p_n,0)$ respectively, as $N\rightarrow\infty$. Then we put the results together to get an approximation to $\tau_{\text{int}}(p_n,0)$. The asymptotic behaviour of $\tau_{\text{int}}(q_n,0)$ as $N\rightarrow\infty$ will be studied likewise.

\subsection{Inner region approximation}
In order to get an asymptotic expression for $\tau_{\text{int}}^{\text{in}}(p_n,0)$ as $N\rightarrow\infty$, we first look for the expansion of each term in the summation appearing in Eq.~\eqref{tau_p_inner}. Given a dislocation associated with the inner region, the distance from its $x$-coordinate (for example, $p_j$ or $q_j$), to $x_n$, the $x$-coordinate of the center of $n\text{-th}$ dislocation pair is small compared to the length of the computational domain (which equals to $1$ after non-dimensionalisation). Hence we use Taylor expansion to asymptotically express $p_j$ and $q_j$ in terms of $x_n$. This is accomplished in two steps: first we relate $p_j$ and $q_j$ to $x_j$ and then we relate $x_j$ to $x_n$. The first step has been achieved by Eq.~\eqref{pqn_from_xnzetan}. For the second step, we re-write Eq.~\eqref{DDPF_def} by
\beq \label{relate_xj_to_xn_inner}
\phi(t,x_j) = \frac{j}{N} = \frac{n}{N} + \frac{j-n}{N} = \phi(t,x_n) + \frac{j-n}{N}.
\eeq
Applying $\phi^{-1}$ to both sides of Eq.~\eqref{relate_xj_to_xn_inner}, noting that $\phi^{-1}(j/N)=x_j$, gives
\beq \label{relate_xj_to_xn_inner0}
x_j = \phi^{-1}\left(\phi(t,x_n) + \frac{j-n}{N}\right).
\eeq
Since $|\frac{j-n}{N}| \le \frac{K}{N} \ll 1$ for all $x_j\in\Omega_{\text{in}}^n$, we expand Eq.~\eqref{relate_xj_to_xn_inner0} in terms of $\frac{j-n}{N}$ to obtain
\beq \label{expan_xj_inner}
x_j \sim x_n + \frac{j-n}{N}\cdot\frac1{\phi'_n} - \frac{(j-n)^2}{N^2}\cdot\frac{\phi_n''}{2(\phi'_n)^3} + \frac{(j-n)^3}{N^3}\cdot\frac{(3(\phi_n'')^2-\phi_n'\phi_n''')}{6(\phi'_n)^5} + \CO\left(\frac{K^4}{N^4}\right).
\eeq
Recall that an index $n$ on $\phi$ or $\zeta$ denotes that the evaluation is made at $x_n$. Using Eq.~\eqref{expan_xj_inner}, we Taylor expand $\zeta_j$ about $x_n$ to give
\beq \label{expan_zetaj_inner}
\zeta_j = \zeta(t,x_j) \sim \zeta_n + \frac1{N}\cdot\frac{(j-n)\zeta_n'}{\phi'_n} + \frac1{N^2}\cdot\frac{(j-n)^2(\zeta_n''\phi_n'-\phi_n''\zeta_n')}{2(\phi'_n)^3} + \CO\left(\frac{K^3}{N^3}\right).
\eeq

Combining Eqs.~\eqref{pqn_from_xnzetan}, \eqref{expan_xj_inner} and \eqref{expan_zetaj_inner}, we asymptotically express $p_j$ near $x_n$ by
\beq \label{expan_pj_inner}
\begin{aligned}
p_j &\sim x_n + \frac1{N}\cdot\left(\frac{j-n}{\phi'_n}-\frac{\zeta_n}{2}\right) - \frac1{2N^2}\cdot\left(\frac{(j-n)\zeta_n'}{\phi'_n} + \frac{(j-n)^2\phi_n''}{(\phi'_n)^3} \right) \\
&\quad + \frac{(j-n)^2}{N^3}\cdot\left(\frac{\zeta_n'\phi_n''}{4(\phi_n')^3} + \frac{(j-n)(\phi_n'')^2}{2(\phi_n')^5} - \frac{\zeta_n''}{4(\phi_n')^2} - \frac{(j-n)\phi_n'''}{6(\phi_n')^4}\right) + \CO\left(\frac{K^4}{N^4}\right).
\end{aligned}
\eeq
Similarly we find
\beq \label{expan_qj_inner}
\begin{aligned}
q_j &\sim x_n + \frac1{N}\cdot\left(\frac{j-n}{\phi'_n}+\frac{\zeta_n}{2}\right) + \frac1{2N^2}\cdot\left(\frac{(j-n)\zeta_n'}{\phi'_n} - \frac{(j-n)^2\phi_n''}{(\phi'_n)^3} \right) \\
&\quad - \frac{(j-n)^2}{N^3}\cdot\left(\frac{\zeta_n'\phi_n''}{4(\phi_n')^3} - \frac{(j-n)(\phi_n'')^2}{2(\phi_n')^5} - \frac{\zeta_n''}{4(\phi_n')^2} + \frac{(j-n)\phi_n'''}{6(\phi_n')^4}\right) + \CO\left(\frac{K^4}{N^4}\right).
\end{aligned}
\eeq

Then incorporating Eqs.~\eqref{expan_pj_inner} and \eqref{expan_qj_inner} into \eqref{tau_p_inner}, we obtain the expansion of $\tau_{\text{int}}^{\text{in}}(p_n,0)$ as
\beq \label{tau_p_est_inner}
\begin{aligned}
\tau_{\text{int}}^{\text{in}}(p_n,0) &\sim  (\pi \phi_n') \cdot G_0(2\pi\zeta_n\phi_n', 2\pi S\phi_n') +\frac{2\zeta_n\phi_n'}{K} - \frac{\zeta_n\phi_n'}{K^2} \\
& \quad - \frac{\phi_n''}{N\phi_n'}\cdot G_{11}(2\pi\zeta_n\phi_n', 2\pi S\phi_n') -\frac{(\zeta_n\phi_n')'}{N}\cdot G_{12}(2\pi\zeta_n\phi_n', 2\pi S\phi_n') \\
& \quad - \frac{\phi_n'\zeta_n'}{N}\cdot G_{13}(2\pi\zeta_n\phi_n', 2\pi S\phi_n') + o\left(\frac1{N}\right),
\end{aligned}
\eeq
where
\beq \label{term_G0}
G_0(\alpha,\beta) = \frac{\sin\alpha}{\cosh\beta-\cos\alpha } - \frac{\beta \sin\alpha\sinh\beta}{(\cos\alpha - \cosh\beta)^2},
\eeq
\beq \label{term_G11}
\begin{aligned}
&G_{11}(\alpha,\beta) = -\frac1{2} - \frac{\alpha\sin\alpha + 2\beta\sinh\beta}{2(\cos\alpha-\cosh\beta)} + \frac{5\beta^2(1-\cos\alpha\cosh\beta)}{4(\cos\alpha-\cosh\beta)^2} - \frac{3\alpha\beta\sin\alpha\sinh\beta}{2(\cos\alpha-\cosh\beta)^2} \\
&\quad + \frac{\beta^3\sinh\beta(1-\cos\alpha\cosh\beta+\sin^2\alpha) }{4(\cos\alpha-\cosh\beta)^3} + \frac{\alpha\beta^2\sin\alpha(1-\cos\alpha\cosh\beta-\sinh^2\beta) }{2(\cos\alpha-\cosh\beta)^3},
\end{aligned}
\eeq
\beq \label{term_G12}
G_{12}(\alpha,\beta) = - \frac{\pi\alpha(1-\cos\alpha\cosh\beta)}{2(\cos\alpha-\cosh\beta)^2} - \frac{\pi\alpha\beta\sinh\beta(1-\cos\alpha\cosh\beta+\sin^2\alpha) }{2(\cos\alpha-\cosh\beta)^3}.
\eeq
and
\beq \label{term_G13}
\begin{aligned}
&G_{13}(\alpha,\beta) =\\ &\quad -\frac{\pi\sin\alpha}{2}\left(\frac1{\cos\alpha-\cosh\beta} + \frac{3\beta\sinh\beta }{(\cos\alpha-\cosh\beta)^2} - \frac{\beta^2(1-\cos\alpha\cosh\beta-\sinh^2\beta)}{(\cos\alpha-\cosh\beta)^3}\right).
\end{aligned}
\eeq
The detailed derivation of Eq.~\eqref{tau_p_est_inner} is given in the supplementary materials. We will find that the internal resolved shear stress components accounting for the pair density evolution arise at $\CO(1/N)$. Thus, unless specified, the expansions to all resolved shear stresses will be truncated at $o(1/N)$. To ensure this accuracy, we further choose $K \sim \sqrt{N}$.

\subsection{Outer region approximation}
For $x_j$ belonging to the outer region, the expansion given by Eq.~\eqref{expan_xj_inner} no longer holds since $(j-n)/N$ can grow as large as $\CO(1)$. However, according to Eq.~\eqref{outer_region_def}, we have
\beq \label{estimate_xjxn_outer}
\frac{K}{N} < |\phi(x_n)-\phi(x_j)| = \phi'(c_0)|x_n-x_j|,
\eeq
where $c_0$ takes some value between $x_j$ and $x_n$.
Eq.~\eqref{estimate_xjxn_outer} suggests that
\beq\label{distance_outer}
|x_j-x_n| \gg 1/N,
\eeq
for all $x_j\in\Omega_{\text{out}}^n$. Eq.~\eqref{distance_outer} implies that the distance between $x_n$ and the centre of a dislocation pair associated with $\Omega^n_{\text{out}}$ is much larger than the spacing between neighbouring dislocation pairs. We obtain the expansion of the resolved shear stress at $(p_n,0)$ due to the $j$-th pair in two steps: first, we relate $p_j$ and $q_j$ to $x_j$, and relate $p_n$ and $q_n$ to $x_n$ by using Eq.~\eqref{pqn_from_xnzetan}; then we expand as $N\rightarrow\infty$ using Eq.~\eqref{distance_outer}.

Following these two steps, we asymptotically expand the resolved shear stress at $(p_n,0)$ due to the $j$-th pair, i.e. the $j$-th term in the summation of Eq.~\eqref{tau_p_outer}, by
\beq \label{tau_jpair_expan_out}
\begin{aligned}
&\frac1{N}\cdot\left(\frac1{p_n-p_j} - \frac{(p_n-q_j)((p_n-q_j)^2-(S/N)^2)}{((p_n -q_j)^2+(S/N)^2)^2}\right) \sim -\frac1{N^2}\cdot \frac{\zeta_j}{(x_n-x_j)^2} \\
& \quad + \frac1{N^3}\cdot \frac{\zeta_n\zeta_j - 3S^2}{(x_n-x_j)^3} + \frac1{N^4}\cdot \frac{3\zeta_j(6S^2-\zeta_n^2)+18S^2\zeta_n- \zeta_j}{4(x_j-x_n)^4} + \CO\left(\frac1{K^5}\right).
\end{aligned}
\eeq
We see from Eq.~\eqref{tau_jpair_expan_out} that the leading-order effect of the stress at $(p_n,0)$ due to the positive dislocation at $(p_j,0)$ cancels with that due to its pair partner located at $(q_n,s)$.

It is worth noting that, when the truncation is made, $1/|x_j-x_n|$ can be as large as $\CO(N/K)$. Besides, the summation made for outer region approximation involves almost $N$ terms. Therefore, to ensure an accuracy of $o(1/N)$ for a resolved stress component, the truncation made at each term in the summation should be at $o(1/N^2)$. This is the reason that a truncation at $\CO(1/K^5)$ is made in Eq.~\eqref{tau_jpair_expan_out}, given $K\sim\sqrt{N}$.

Incorporating Eq.~\eqref{tau_jpair_expan_out} into \eqref{tau_p_outer} gives the expansion of $\tau_{\text{int}}^{\text{out}}(p_n,0)$:
\beq \label{tau_p_expan_out}
\begin{aligned}
\tau_{\text{int}}^{\text{out}}(p_n,0) & \sim -\frac1{N^2}\cdot\sum_{\substack{0\le j<n-K\\n+K < j\le N}}\frac{\zeta_j}{(x_n-x_j)^2} + \frac1{N^3}\cdot \sum_{\substack{0\le j<n-K\\n+K < j\le N}}\frac{\zeta_n\zeta_j - 3S^2}{(x_n-x_j)^3} \\
& \quad + \frac1{N^4}\cdot \sum_{\substack{0\le j<n-K\\n+K < j\le N}} \frac{3\zeta_j(6S^2-\zeta_n^2)+18S^2\zeta_n-\zeta_j}{4(x_j-x_n)^4} + \CO\left(\frac{N}{K^5}\right)
\end{aligned}
\eeq

To evaluate the summations appearing in Eq.~\eqref{tau_p_expan_out}, we make use of the Euler-Maclaurin formula. The details are in the supplementary materials, and the result is
\beq \label{tau_p_est_out}
\begin{aligned}
\tau_{\text{int}}^{\text{out}}(p_n,0) &\sim -\frac{2}{K}\cdot(\zeta_n\phi_n') + \frac1{K^2}\cdot\zeta_n\phi_n'+ \frac1{N}\cdot\left(\frac{\phi_0'\zeta_0}{x_n-x_0} - \frac{\phi_N'\zeta_N}{x_n-x_N}\right) \\
& \quad + \frac1{N}\dashint_{x_0}^{x_N} \frac{(\phi'(a)\zeta(a))'\d a}{x_n-a} + o\left(\frac1{N}\right),
\end{aligned}
\eeq
where the integral is evaluated in the sense of principal value.

\subsection{Total resolved shear stress\label{Sec_total_resolved_shear_stress}}
Now we put the results from the inner expansion by Eq.~\eqref{tau_p_est_inner} and from the outer expansion by \eqref{tau_p_est_out} together to obtain the expansion of $\tau_{\text{int}}(p_n,0)$ as
\beq \label{tau_p_est_n}
\begin{aligned}
&\tau_{\text{int}}(p_n,0) \sim \left(\pi\phi_n'\right)\cdot G_0(2\pi \phi_n'\zeta_n, 2\pi S\phi_n') + \frac1{N}\cdot\left(\frac{\phi_0'\zeta_0}{x_n-x_0} - \frac{\phi_N'\zeta_N}{x_n-x_N}\right) \\
&\quad + \frac1{N}\dashint_{x_0}^{x_N} \frac{(\phi'(a)\zeta(a))'\d a}{x_n-a} - \frac{\phi_n''}{N\phi_n'}\cdot G_{11}(2\pi\phi_n'\zeta_n,2\pi S\phi_n') \\
&\quad - \frac{(\phi_n'\zeta_n)'}{N}\cdot G_{12}(2\pi\phi_n'\zeta_n,2\pi S\phi_n')  - \frac{\phi_n'\zeta_n'}{N}\cdot G_{13}(2\pi\phi_n'\zeta_n,2\pi S\phi_n') + o\left(\frac1{N}\right).
\end{aligned}
\eeq
It is worth noting that the $\CO(1/K)$ and $\CO(1/K^2)$ terms from the inner expansion cancel with their counterparts from the outer expansion. As a result, no trace of the intermediate parameter $K$ is seen in Eq.~\eqref{tau_p_est_n}.

The external stress $\tau_{\text{ext}}$ at $(p_n,0)$ can also be expanded near $(x_n,0)$
\beq \label{tau_ext_p_expan}
\tau_{\text{ext}}(p_n,0) \sim \tau^0_{\text{ext}}(x_n) - \frac{\zeta_n}{2N}\cdot\pd{\tau^0_{\text{ext}}(x_n)}{x} + \CO\left(\frac1{N^2}\right),
\eeq
where for ease of notation we have written $
\tau^0_{\text{ext}}(x) = \tau_{\text{ext}}(x,0)$ and $\pd{\tau^0_{\text{ext}}(x)}{x} = \pd{\tau_{\text{ext}}(x,0)}{x}$.
Similarly,
\beq \label{tau_ext_q_expan}
\tau_{\text{ext}}(q_n,s) \sim \tau^0_{\text{ext}}(x_n) + \frac1{N}\left(\frac{\zeta_n}{2}\cdot \pd{\tau^0_{\text{ext}}(x_n)}{x} + S\cdot\pd{\tau^0_{\text{ext}}(x_n)}{y}\right)+ \CO\left(\frac1{N^2}\right),
\eeq
where $
\pd{\tau^0_{\text{ext}}(x)}{y} = \pd{\tau_{\text{ext}}(x,0)}{y}$.
Here $\pd{\tau^0_{\text{ext}}}{x}$ and $\pd{\tau^0_{\text{ext}}}{y}$ are the stress gradients, which capture the difference in the externally applied stress field evaluated at each component of a dipole pair.

Thus the (non-dimensional) total resolved shear stress at $(p_n,0)$ is given by
\beq \label{tau_total_p}
\begin{aligned}
&\tau_{\text{tot}}(p_n,0) \sim \left(\pi\phi_n'\right)\cdot G_0(2\pi \phi_n'\zeta_n, 2\pi S\phi_n')+ \tau^0_{\text{ext}}(x_n) \\
&\quad + \frac1{N}\cdot\left(\frac{\phi_0'\zeta_0}{x_n-x_0} - \frac{\phi_N'\zeta_N}{x_n-x_N}\right) + \frac1{N}\dashint_{x_0}^{x_N} \frac{(\phi'(a)\zeta(a))'\d a}{x_n-a}\\
&\quad - \frac{\phi_n''}{N\phi_n'}\cdot G_{11}(2\pi\phi_n'\zeta_n,2\pi S\phi_n') - \frac{(\phi_n'\zeta_n)'}{N}\cdot G_{12}(2\pi\phi_n'\zeta_n,2\pi S\phi_n') \\
&\quad - \frac{\phi_n'\zeta_n'}{N}\cdot G_{13}(2\pi\phi_n'\zeta_n,2\pi S\phi_n')- \frac{\zeta_n}{2N} \pd{\tau^0_{\text{ext}}(x_n)}{x} + o\left(\frac1{N}\right),
\end{aligned}
\eeq
where we recall that $G_0$, $G_{11}$, $G_{12}$ and $G_{13}$ are defined by Eqs.~\eqref{term_G0} - \eqref{term_G13}. Similarly,
\beq \label{tau_total_q}
\begin{aligned}
&\tau_{\text{tot}}(q_n,s) \sim \left(\pi\phi_n'\right)\cdot G_0(2\pi \phi_n'\zeta_n, 2\pi S\phi_n')+ \tau^0_{\text{ext}}(x_n) \\
&\quad + \frac1{N}\cdot\left(\frac{\phi_0'\zeta_0}{x_n-x_0} - \frac{\phi_N'\zeta_N}{x_n-x_N}\right) + \frac1{N}\dashint_{x_0}^{x_N} \frac{(\phi'(a)\zeta(a))'\d a}{x_n-a}\\
&\quad + \frac{\phi_n''}{N\phi_n'}\cdot G_{11}(2\pi\phi_n'\zeta_n,2\pi S\phi_n') + \frac{(\phi_n'\zeta_n)'}{N}\cdot G_{12}(2\pi\phi_n'\zeta_n,2\pi S\phi_n') \\
&\quad + \frac{\phi_n'\zeta_n'}{N}\cdot G_{13}(2\pi\phi_n'\zeta_n,2\pi S\phi_n') + \frac{\zeta_n}{2N}\cdot \pd{\tau^0_{\text{ext}}(x_n)}{x} + \frac{S}{N}\cdot\pd{\tau^0_{\text{ext}}(x_n)}{y} + o\left(\frac1{N}\right).
\end{aligned}
\eeq

\section{Dislocation dynamical model at the continuum level\label{Sec_continuum_model}}
In this section, we derive the governing equations for the two field variables $\phi$ and $\zeta$ at the continuum level. We first consider the continuous description for the equilibrium states, where the total resolved shear stress at each dislocation vanishes.

\subsection{Governing equations for the equilibrium states}
When all dipoles are in equilibrium, the total resolved shear stress $\tau_{\text{tot}}(p_n,0)$ and $\tau_{\text{tot}}(q_n,s)$ should be zero for all $n$ according to the laws of motion \eqref{mobility_law_pos_dimensionless} and \eqref{mobility_law_neg_dimensionless}. It is worth noting that since the resulting equations are established in an asymptotic sense, we also need to expand $\phi$ and $\zeta$ as
\beq \label{phi_expan}
\phi(t,x) \sim \phi^{(0)}(t,x) + \frac{\phi^{(1)}(t,x)}{N} + \cdots.
\eeq
and
\beq \label{zeta_expan}
\zeta(t,x) \sim \zeta^{(0)}(t,x) + \frac{\zeta^{(1)}(t,x)}{N} + \cdots,
\eeq
respectively. Substituting the above expansions into Eqs.~\eqref{tau_total_p} and \eqref{tau_total_q} gives
\beq \label{tau_total_pn}
\begin{aligned}
\tau_{\text{tot}}(p_n,0) &\sim  \left(\pi(\phi_n^{(0)})'\right)\cdot G_0(2\pi (\phi_n^{(0)})'\zeta_n^{(0)}, 2\pi S(\phi_n^{(0)})')+ \tau^0_{\text{ext}}(x_n) \\
&\quad + \frac1{N}\cdot \left(\tau_a^n - \tau_b^n - \frac{\zeta_n^{(0)}}{2}\cdot \pd{\tau^0_{\text{ext}}(x_n)}{x}\right)  + o\left(\frac1{N}\right)
\end{aligned}
\eeq
and
\beq \label{tau_total_qn}
\begin{aligned}
\tau_{\text{tot}}(q_n,s) &\sim  \left(\pi(\phi_n^{(0)})'\right)\cdot G_0(2\pi (\phi_n^{(0)})'\zeta_n^{(0)}, 2\pi S(\phi_n^{(0)})')+ \tau^0_{\text{ext}}(x_n) \\
&\quad + \frac1{N}\cdot \left(\tau_a^n + \tau_b^n + \frac{\zeta_n^{(0)}}{2}\cdot \pd{\tau^0_{\text{ext}}(x_n)}{x} + S\cdot \pd{\tau^0_{\text{ext}}(x_n)}{x}\right)  + o\left(\frac1{N}\right),
\end{aligned}
\eeq
respectively, where
\beq \label{tau_int_O1a}
\begin{aligned}
\tau_a^n &= \frac{(\phi_0^{(0)})'\zeta_0^{(0)}}{x_n-x_0} - \frac{(\phi_N^{(0)})'\zeta_N^{(0)}}{x_n-x_N} + \dashint_{x_0}^{x_N} \frac{(\phi'(a)\zeta(a)^{(0)})'\d t}{x_n-t}\\
&\quad + \zeta_n^{(1)} \pd{G_0(2\pi (\phi_n^{(0)})'\zeta_n^{(0)}, 2\pi S(\phi_n^{(0)})')}{\zeta_n^{(0)}} + (\phi_n^{(1)})' \pd{G_0(2\pi (\phi_n^{(0)})'\zeta_n^{(0)}, 2\pi S(\phi_n^{(0)})')}{ (\phi_n^{(0)})'},
\end{aligned}
\eeq
\beq \label{tau_int_O1b}
\begin{aligned}
\tau_b^n &= \frac{(\phi_n^{(0)})''}{(\phi_n^{(0)})'}\cdot G_{11}(2\pi(\phi_n^{(0)})'\zeta_n^{(0)},2\pi S(\phi_n^{(0)})') \\
&\quad + ((\phi_n^{(0)})'\zeta_n^{(0)})'\cdot G_{12}(2\pi(\phi_n^{(0)})'\zeta_n^{(0)},2\pi S(\phi_n^{(0)})') \\
&\quad + (\phi_n^{(0)})'(\zeta_n^{(0)})'\cdot G_{13}(2\pi(\phi_n^{(0)})'\zeta_n^{(0)},2\pi S(\phi_n^{(0)})').
\end{aligned}
\eeq

Now letting the right hand side of Eqs.~\eqref{tau_total_pn} and \eqref{tau_total_qn} vanish and equating coefficients of the same powers of $N$, we obtain at leading order,
\beq \label{eqn_fb_general_n1}
\pi(\phi_n^{(0)})'\cdot G_0(2\pi (\phi_n^{(0)})'\zeta_n^{(0)}, 2\pi S(\phi_n^{(0)})') + \tau^0_{\text{ext}}(x_n) = 0.
\eeq
There are $N+1$ equations for the $2(N+1)$ unknowns $\{\phi_n^{(0)}\}_{n=0}^N$ and $\{\zeta_n^{(0)}\}_{n=0}^N$.

To close the system, we need to proceed to higher order in the expansion. At $\CO(1/N)$, we find
\beq \label{eqn_fb_general_n2a}
\tau_a^n - \tau_b^n - \frac{\zeta_n^{(0)}}{2}\cdot \pd{\tau^0_{\text{ext}}(x_n)}{x} = 0
\eeq
and
\beq \label{eqn_fb_general_n2b}
\tau_a^n + \tau_b^n + \frac{\zeta_n^{(0)}}{2}\cdot \pd{\tau^0_{\text{ext}}(x_n)}{x} + S\cdot \pd{\tau^0_{\text{ext}}(x_n)}{x} = 0.
\eeq
Now there are $4(N+1)$ unknowns $\{\phi_n^{(0)}\}_{n=0}^N$, $\{\phi_n^{(1)}\}_{n=0}^N$, $\{\zeta_n^{(0)}\}_{n=0}^N$ and $\{\zeta_n^{(1)}\}_{n=0}^N$ and $3(N+1)$ equations. However, we subtract Eq.~\eqref{eqn_fb_general_n2a} from \eqref{eqn_fb_general_n2b} to eliminate $\zeta_n^{(1)}$ and $\phi_n^{(1)}$, both of which only appear in $\tau_a^n$, to give
\beq \label{eqn_fb_general_n2}
2\tau_b^n + \zeta_n^{(0)}\cdot \pd{\tau^0_{\text{ext}}(x_n)}{x} + S \cdot \pd{\tau^0_{\text{ext}}(x_n)}{y} = 0.
\eeq
Eqs.~\eqref{eqn_fb_general_n1} and \eqref{eqn_fb_general_n2} form a system consisting of $2(N+1)$ equations for $2(N+1)$ unknowns ($\{\phi_n^{(0)}\}_{n=0}^N$ and $\{\zeta_n^{(0)}\}_{n=0}^N$). Henceforth we drop the superscript ``$^{(0)}$'', because only the leading-order effects are taken into account.

Since $x_n$ is densely distributed in the domain, we rewrite our equations valid at every $x_n$ as equations valid for all $x$. Therefore, we drop the index $n$ and re-write Eqs.~\eqref{eqn_fb_general_n1} and \eqref{eqn_fb_general_n2} as
\begin{subequations}
\beq\label{eqn_fb_general1}
\frac{\pi\phi'\sin(2\pi\phi'\zeta)}{\cosh(2\pi S\phi')-\cos(2\pi\phi'\zeta)}\cdot\left(1-\frac{2\pi S\phi' \sinh(2\pi S\phi')}{\cosh(2\pi S\phi')-\cos(2\pi\phi'\zeta)}\right) + \tau_{\text{ext}}^0 = 0
\eeq
\beq \label{eqn_fb_general2}
\begin{aligned}
0 & = \frac{2\phi''}{\phi'}\cdot G_{11}(2\pi\phi'\zeta,2\pi S\phi') + 2(\phi'\zeta)'\cdot G_{12}(2\pi\phi'\zeta,2\pi S\phi') \\
&\quad + 2(\phi'\zeta')\cdot G_{13}(2\pi\phi'\zeta,2\pi S\phi') + \zeta\frac{\partial \tau_{\text{ext}}^0}{\partial x} + S\frac{\partial \tau_{\text{ext}}^0}{\partial y},
\end{aligned}
\eeq
\end{subequations}
respectively, where we recall that $G_{11}$, $G_{12}$ and $G_{13}$ are defined by Eqs.~\eqref{term_G11} - \eqref{term_G13}. Eqs.~\eqref{eqn_fb_general1} and \eqref{eqn_fb_general2} are the two equations for the two field variables $\phi$ and $\zeta$ derived at the continuum level when the row of dipoles rest in their equilibrium states. It is worth noting that Eq.~\eqref{eqn_fb_general1} comes from the leading-order force balance and Eq.~\eqref{eqn_fb_general2} comes from the difference in the force balance equations obtained at the next order.

\subsection{Governing equations for the dynamics\label{Sec_dynamics}}
Now we consider reformulating the discrete dislocation dynamics governed by Eqs.~\eqref{mobility_law_pos_dimensionless} to \eqref{tau_internal_neg_n_dimensionless} at the continuum level by looking for evolution equations for $\phi$ and $\zeta$.

We know by definition that $\phi(t,x_n(t))=n/N$ at any time $t$. Taking the derivative with respect to $t$ on both sides gives
\beq \label{eqn_phi_evo0}
\frac{\partial \phi_n}{\partial t} + \frac{\d x_n}{\d t}\cdot\frac{\partial \phi_n}{\partial x} = 0.
\eeq
According to the definition of $x_n$ given by Eq.~\eqref{xn_def}, we have
\beq \label{dxdt_exp}
\frac{\d x_n}{\d t} = \frac{\tau_{\text{tot}}(p_n,0) -\tau_{\text{tot}}(q_n,s)}{2}
\eeq
where the laws of motion~\eqref{mobility_law_pos_dimensionless} and \eqref{mobility_law_neg_dimensionless} are employed.
With the asymptotic expansions for $\tau_{\text{tot}}(p_n,0)$ and $\tau_{\text{tot}}(q_n,s)$ given by Eqs~\eqref{tau_total_p} and \eqref{tau_total_q}, respectively, we incorporate Eq.~\eqref{dxdt_exp} into \eqref{eqn_phi_evo0} to get
\beq \label{eqn_phi_evo_n}
\begin{aligned}
&\frac{\partial \phi_n}{\partial t} - \frac1{N}\left(\frac{\phi_n''}{\phi_n'}G_{11}(2\pi\phi_n'\zeta_n,2\pi S\phi_n') + (\phi_n'\zeta_n)'G_{12}(2\pi\phi_n'\zeta_n,2\pi S\phi_n')\right) \frac{\partial \phi_n}{\partial x} \\
&\quad - \frac1{N}\left((\phi_n'\zeta_n')G_{13}(2\pi\phi_n'\zeta_n,2\pi S\phi_n') + \frac{\zeta}{2}\frac{\partial \tau_{\text{ext}}^0(x_n)}{\partial x} + \frac{S}{2}\frac{\partial \tau_{\text{ext}}^0(x_n)}{\partial y}\right) \frac{\partial \phi_n}{\partial x} \sim o\left(\frac1{N}\right).
\end{aligned}
\eeq

Again we drop the subscript $n$ to rewrite Eq.~\eqref{eqn_phi_evo_n} as a differential equation valid for all $x$ by
\beq \label{eqn_phi_evo}
\frac{\partial \phi}{\partial t}  - \frac1{N}\left(\tau_b + \frac{\zeta}{2}\cdot \pd{\tau^0_{\text{ext}}}{x} + \frac{S}{2} \cdot \pd{\tau^0_{\text{ext}}}{y}\right)\cdot\frac{\partial \phi}{\partial x} \sim o\left(\frac1{N}\right),
\eeq
where
\beq \label{tau_b}
\tau_b = \frac{\phi''}{\phi'}G_{11}(2\pi\phi'\zeta,2\pi S\phi') + (\phi'\zeta)'G_{12}(2\pi\phi'\zeta,2\pi S\phi') + \phi'\zeta' G_{13}(2\pi\phi'\zeta,2\pi S\phi').
\eeq
Eq.~\eqref{eqn_phi_evo} can be considered as the evolution equation for $\phi$.

It can be seen from Eq.~\eqref{eqn_phi_evo} that the evolution speed of $\phi$ is as small as $\CO(1/N)$. This suggests that the natural time scale associated with the evolution of $\phi$, the dislocation pair density potential, is characterised by a slow-varying temporal variable given by $t_{\text{s}} = Nt$. Eq.~\eqref{eqn_phi_evo} then gives at leading order
\beq \label{eqn_phi_evo_ts0}
\frac{\partial \phi}{\partial t_{\text{s}}} - \left(\tau_b + \frac{\zeta}{2} \cdot \pd{\tau^0_{\text{ext}}}{x} + \frac{S}{2}\cdot \pd{\tau^0_{\text{ext}}}{y}\right)\cdot\frac{\partial \phi}{\partial x} = 0.
\eeq

On the other hand, according to the definition of $\zeta$ by Eq.~\eqref{zeta_def}, we have
\beq\label{eqn_zetan_evo0}
\frac{\partial \zeta_n}{\partial t} + \frac{\d x_n}{\d t}\cdot\frac{\partial \zeta_n}{\partial x} = \frac{\d \zeta_n(t,x(t))}{\d t} = N\cdot\left(\frac{\d q_n}{\d t} - \frac{\d p_n}{\d t}\right).
\eeq
Combining Eqs.~\eqref{mobility_law_pos_dimensionless}, \eqref{mobility_law_neg_dimensionless}, \eqref{tau_total_p}, \eqref{tau_total_q}, \eqref{dxdt_exp} and \eqref{eqn_zetan_evo0} then dropping the subscript $n$, we find
\beq\label{eqn_zeta_evo}
\frac{\partial \zeta}{\partial t} \sim \frac{2N\pi\phi'\sin(2\pi\phi'\zeta)}{\cosh(2\pi S\phi')-\cos(2\pi\phi'\zeta)}\left(\frac{2\pi S\phi' \sinh(2\pi S\phi')}{\cosh(2\pi S\phi')-\cos(2\pi\phi'\zeta)}-1\right) - 2N\tau_{\text{ext}}^0 + \CO(1).
\eeq
It is seen from Eq.~\eqref{eqn_zeta_evo} that $\zeta$ evolves as fast as $\CO(N)$. This means $\zeta$ should be studied at a fast temporal scale characterised by $
t_{\text{f}} = t/N$.
Then the leading-order equation for $\zeta$ is
\beq \label{eqn_zeta_evo_tf}
\frac{\partial \zeta}{\partial t_{\text{f}}} = - \frac{2\pi\phi'\sin(2\pi\phi'\zeta)}{\cosh(2\pi S\phi')-\cos(2\pi\phi'\zeta)}\cdot\left(1 -\frac{2\pi S\phi' \sinh(2\pi S\phi')}{\cosh(2\pi S\phi')-\cos(2\pi\phi'\zeta)}\right) - 2\tau_{\text{ext}}^0.
\eeq
A comparison between Eqs. \eqref{eqn_phi_evo_ts0} and \eqref{eqn_zeta_evo_tf} shows that the evolution of $\zeta$ is much faster than that of $\phi$. Hence $\zeta$ can be considered varying quasi-statically on the time scale characterised by $t_{\text{s}}$, on which $\phi$ naturally evolves, provided stable equilibria exist for Eq.~\eqref{eqn_zeta_evo_tf}. In fact, Eq.~\eqref{eqn_zeta_evo_tf} can be written by $\frac{\partial \zeta}{\partial t_{\text{f}}}=-\frac{\partial \CF}{\partial \zeta}$, where $\CF$ is the generalised free energy density with respect to $\zeta$, given by
\beq \label{free_energy}
\CF = \log\left(\cosh(2\pi S\phi')-\cos(2\pi \zeta\phi')\right) + \frac{2\pi\phi'S\sinh(2\pi S\phi')}{\cosh(2\pi S\phi')-\cos(2\pi \zeta\phi')} + 2\zeta\tau_{\text{ext}}^0.
\eeq
Since $\phi'$ is assumed static on the fast scale, the stable equilibria of Eq.~\eqref{eqn_zeta_evo_tf} are identified wherever $\CF$ attains its local minimum with respect to $\zeta$. It will be shown numerically later that given $\phi'$ and $S$, stable equilibria exist for Eq.~\eqref{eqn_zeta_evo_tf} when $|\tau_{\text{ext}}^0|$ falls below some critical value.

Therefore, the dynamics of a row of dislocation dipoles at the continuum level can be described by the following coupled equations:
\begin{subequations}
\beq \label{eqn_zeta_static}
\frac{\pi\phi'\sin(2\pi\phi'\zeta)}{\cosh(2\pi S\phi')-\cos(2\pi\phi'\zeta)}\cdot\left(1 -\frac{2\pi S\phi' \sinh(2\pi S\phi')}{\cosh(2\pi S\phi')-\cos(2\pi\phi'\zeta)}\right) + \tau_{\text{ext}}^0 = 0,
\eeq
\beq \label{eqn_phi_evo_ts}
\frac{\partial \phi}{\partial t_{\text{s}}} - \left(\tau_b + \frac{\zeta}{2} \cdot \pd{\tau^0_{\text{ext}}}{x} + \frac{S}{2}\cdot \pd{\tau^0_{\text{ext}}}{y}\right)\cdot\frac{\partial \phi}{\partial x} = 0,
\eeq
\end{subequations}
where $\tau_b$ was defined by Eq.~\eqref{tau_b}, provided the stable equilibria of Eq.~\eqref{eqn_zeta_static} exist. Noted that Eq.~\eqref{eqn_zeta_static} is effectively the leading order force balance equation \eqref{eqn_fb_general1}.

\section{Equilibria at the continuum level\label{Sec_equilibria}}
In this section, we will analyse the equilibrium states at the continuum level determined by Eqs.~\eqref{eqn_fb_general1} and \eqref{eqn_fb_general2}. We will begin with the case where the externally applied stress vanishes on $y=0$. In this case, two types of possibly stable configurations are found as a result of the leading-order force balance equation and a natural transition between different equilibrium patterns due to instability is seen. At the next order, the detailed equations for $\phi$ and $\zeta$ corresponding to various equilibrium states will be derived. The analytical results will then be validated through comparison with the numerical solutions to the same problem by using the DDD model. In the end of this section, we will analyse the equilibrium under arbitrary externally-applied stresses.

\subsection{Equilibria under an external stress field which vanishes on $y=0$}
We now analyse Eq.~\eqref{eqn_fb_general1} and \eqref{eqn_fb_general2} by starting with a simple case where the externally applied resolved shear stress vanishes on $y=0$, i.e. $\tau^0_{\text{ext}}=0$. Note that the stress gradient $\partial \tau_{\text{ext}}^0/\partial y$ need not vanish.

\subsubsection{Implication from the leading-order force balance equation\label{Sec_fb_leading_order_zero}}
When $\tau^0_{\text{ext}}=0$, the leading-order force balance equation \eqref{eqn_fb_general1} becomes
\beq \label{eqn_zeta_detail_ext0}
\frac{\sin(2\pi\phi'\zeta)}{\cosh(2\pi S\phi')-\cos(2\pi\phi'\zeta)}\cdot\left(1-\frac{2\pi S\phi' \sinh(2\pi S\phi')}{\cosh(2\pi S\phi')-\cos(2\pi\phi'\zeta)}\right) = 0.
\eeq
Eq.~\eqref{eqn_zeta_detail_ext0} can be regarded as an implicit relation between the two quantities $\phi'\zeta$ and $\phi'S$. In fact, these two quantities are physically meaningful. Since the pair density $\phi'$ can be approximated by the reciprocal of the spacing between two neighbouring pair centers scaled by $N$, and $S$ is the slip plane gap rescaled by $N$, $\phi'S$ captures the ratio of slip plane gap to the inter-spacing of neighbouring pairs. Also since $\zeta/N$ measures the pair width at $x$ according to Eq.~\eqref{zeta_def}, $\phi'\zeta$ measures the ratio of the pair width to the spacing of neighbouring pairs.

\begin{figure}[!ht]
  \centering
  \subfigure[Type I]{\includegraphics[width=.25\textwidth]{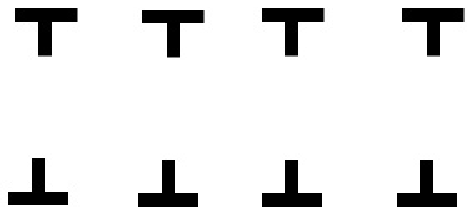}}
  \subfigure[Type II]{\includegraphics[width=.25\textwidth]{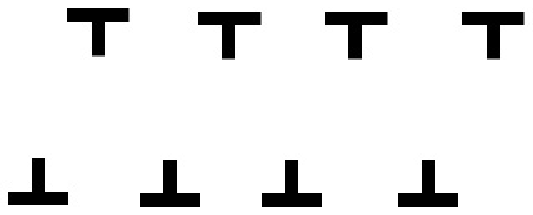}}
  \subfigure[Type III]{\includegraphics[width=.25\textwidth]{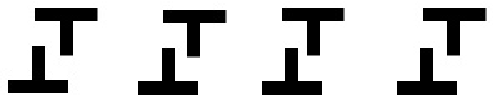}}
  \caption{Three types of equilibria: (a) $\zeta=0$; (b) $\zeta\phi'=1/2$ with non-localised structures formed; (c) $\zeta$ satisfies Eq.~\eqref{equilibrium_type3} and localised structures are formed. \label{fig_illu_types}}
\end{figure}

From Eq.~\eqref{eqn_zeta_detail_ext0}, there are three possible choices for $\zeta$ as a function of $\phi'$ and other parameters.
\begin{itemize}
\item Equilibrium Type I when $\zeta=0$. The dislocation substructure is shown in Fig.~\ref{fig_illu_types}(a). Within each dislocation pair, the positive and the negative dislocations are vertically aligned.
\item Equilibrium Type II when $\phi'\zeta = 1/2$. The dislocation substructure is shown in Fig.~\ref{fig_illu_types}(b). Since $\phi'\zeta$ represents the ratio of pair width to pair center spacing, $\phi'\zeta = 1/2$ suggests that every negative dislocation lies roughly in the middle of its two neighbouring positive dislocations. We term the equilibrium of this type as a ``non-localised'' structure, because each dislocation is ``shared'' by its two neighbours.
\item Equilibrium Type III when
\beq \label{equilibrium_type3}
\zeta = \frac1{2\pi\phi'}\cos^{-1}\left(\cosh(2\pi S\phi') - 2\pi S\phi' \sinh(2\pi S\phi')\right).
\eeq
The dislocation substructure is shown in Fig~\ref{fig_illu_types}(c). A positive dislocation here is bonded with a negative one to form a real dipole, and the equilibrium of this type is named as a ``localised structure''.

It is worth noting that Eq.~\eqref{equilibrium_type3} only holds when
\beq \label{existence_condition_case3a}
-1 \le \cosh(2\pi S\phi')-2\pi S\phi'\sinh(2\pi S\phi') \le 1,
\eeq
which numerically gives rise to a range for $S\phi'$:
\beq \label{existence_condition_case3}
0 \le S\phi' \le 0.2465.
\eeq
Hence the emergence of Equilibrium Type III is conditional.
\end{itemize}

If we set $X=\phi'\zeta$ and $Y=\phi'S$, the configuration is equivalent to a row of dipoles periodic in $X$, which have been studied in \cite{Zhu_2Ddipoles2014}. Thus the conclusion regarding the stability of the obtained three types of equilibria can be drawn by employing the same arguments proposed by \cite{Zhu_2Ddipoles2014}:
\begin{itemize}
\item Equilibrium Type I ($\zeta=0$) is always unstable.
\item Equilibrium Type II ($\phi'\zeta=1/2$) is only stable when Equilibrium Type III does not exist.
\item Equilibrium Type III ($\zeta$ satisfies Eq.~\eqref{equilibrium_type3}) is always stable as long as it exists.
\end{itemize}

\begin{figure}[!ht]
  \centering
  \subfigure[$\phi'S=0.4 $]{\includegraphics[width=.45\textwidth]{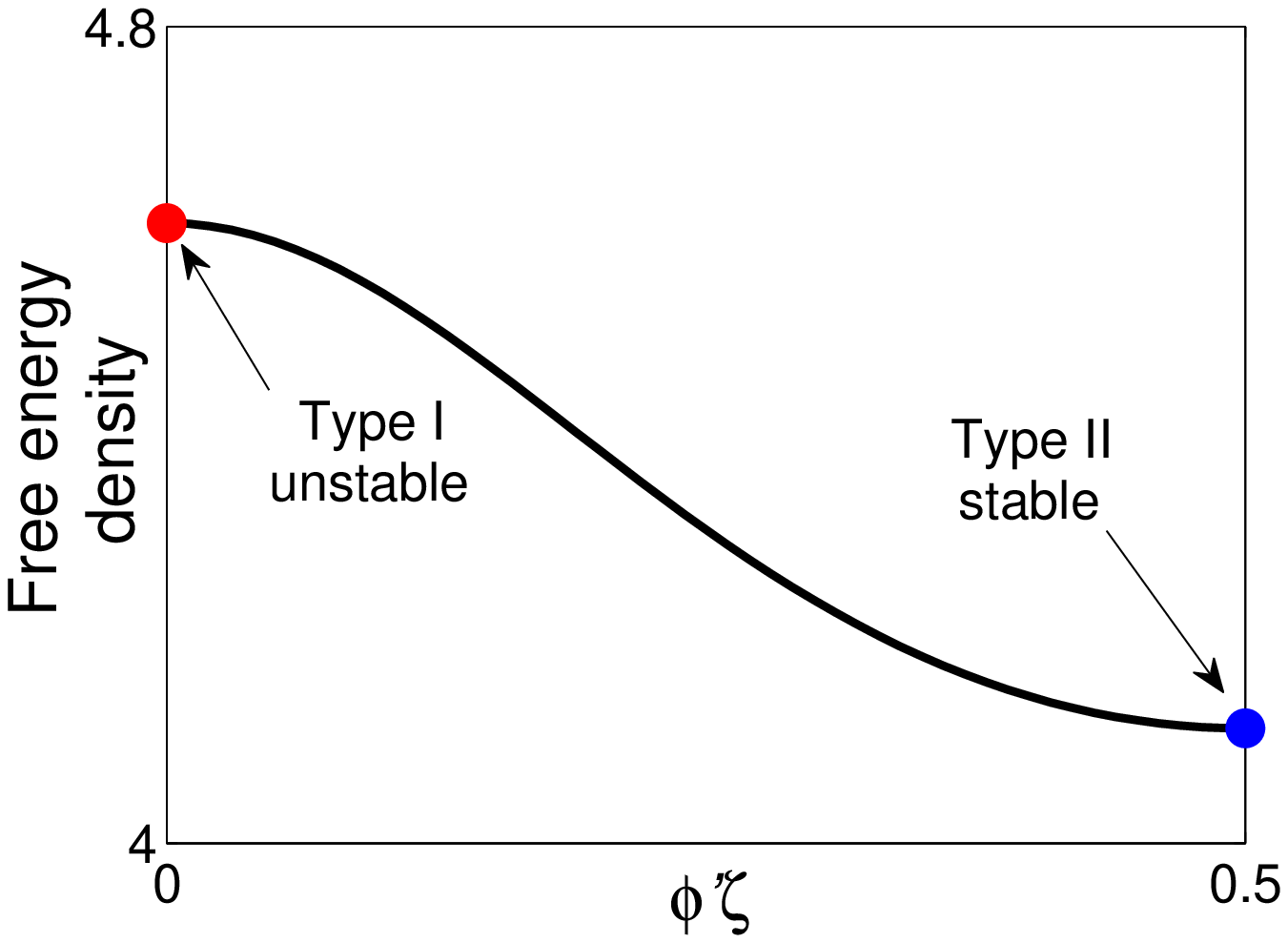}}
  \subfigure[$\phi'S=0.2 $]{\includegraphics[width=.45\textwidth]{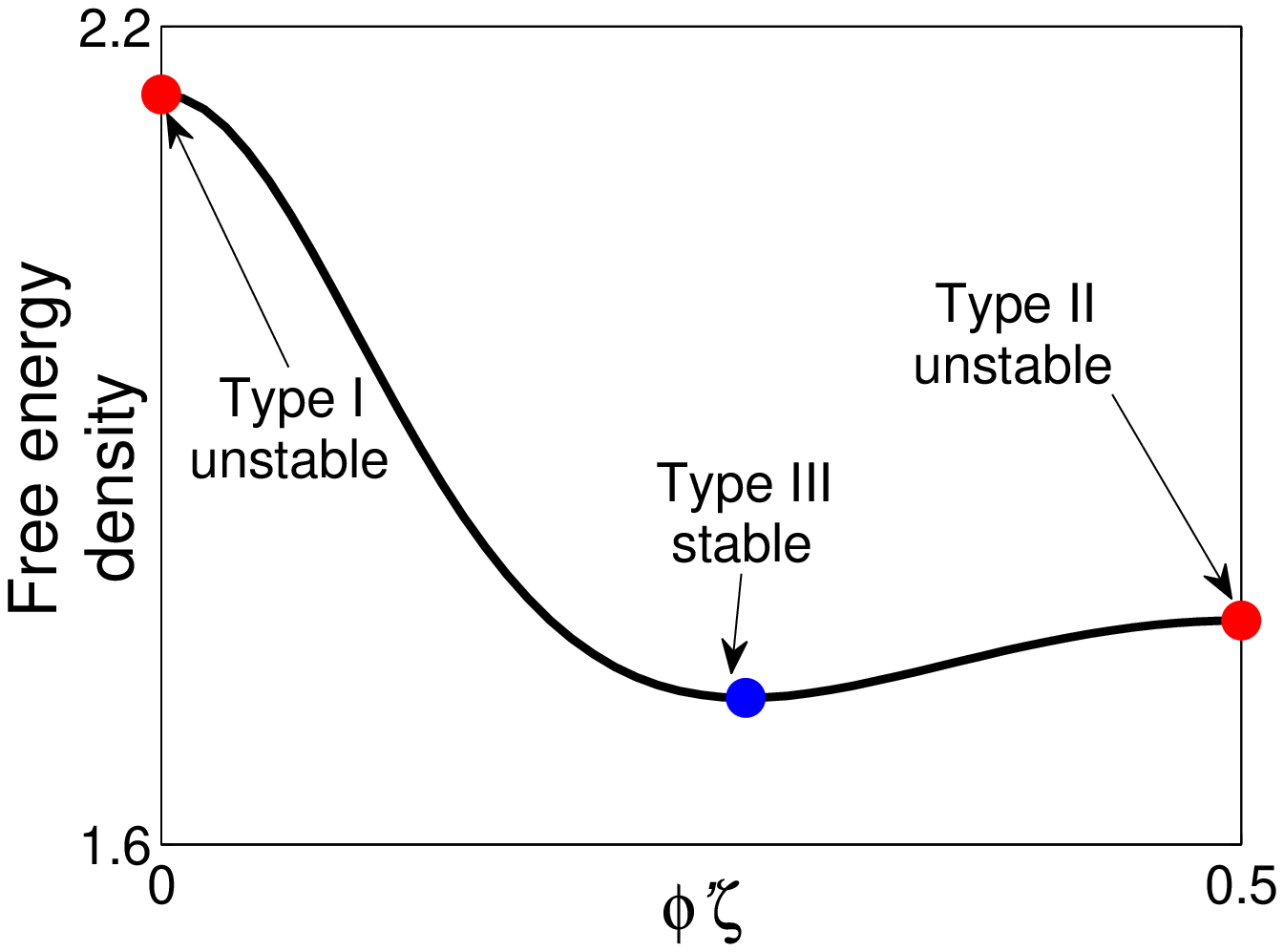}}
  \caption{A stable equilibrium state should correspond to a local minimum of the generalised free energy density $\CF$ given by Eq.~\eqref{free_energy} and $\tau_{\text{ext}}=0$ with respect to $\zeta$. (a) If $\phi'S$ is larger than 0.2465, only two types of equilibria exist and Type II is the stable configuration. (b) If $0<\phi'S<0.2465$, a transition in stability from Type II to Type III takes place. \label{fig_energy_stability}}
\end{figure}

Another way to investigate the stability of the obtained equilibrium states is to look for the local minimum of the free energy density $\CF$ with respect to $\zeta$. When $\tau_{\text{ext}}=0$, $\CF$ given by Eq.~\eqref{free_energy}
are drawn against $\zeta$ for different $\phi'S$ as shown in Fig.~\ref{fig_energy_stability}.
It is seen from Fig.~\ref{fig_energy_stability}(a) that when condition \eqref{existence_condition_case3} is not satisfied, there are two equilibrium states, and Equilibrium Type II is the stable one. When condition \eqref{existence_condition_case3} is met, we have three equilibrium states as shown in Fig.~\ref{fig_energy_stability}, and Equilibrium Type III is the stable one.

Here a natural transition from a non-localised structure (Type II) to a localised structure (Type III) takes place as the slip plane spacing gets narrower or equivalently, as the pair density decreases. Such a transition may be indicative of the formation of the persistent slip bands; further discussion on this issue will be made in \S\ref{Sec_discussion_PSBs}.

\subsubsection{First-order force balance equation for Equilibrium Type II\label{Sec_type2}}
Based on the solutions to the leading-order equation \eqref{eqn_fb_general1}, we now investigate the first-order force balance equation \eqref{eqn_fb_general2}. Here only stable configurations, i.e. Equilibrium Type II and III, are considered.

When $\phi'\zeta = 1/2$ (Type II), one can make use of the fact that $(\phi'\zeta)'=0$ and $\sin(2\pi \phi'\zeta)=0$. This suggests that the terms associated with $G_{12}$ and $G_{13}$ in Eq.~\eqref{eqn_fb_general2} both vanish. Therefore, the equation for $\phi'$ can be obtained as
\beq \label{eqn_phi_case2}
\begin{aligned}
0 &= \frac{2\phi''}{\phi'}\cdot G_{11}(\pi,2\pi S\phi') + S\frac{\partial \tau_{\text{ext}}^0}{\partial y}\\
& = -\frac{\phi''}{\phi'} - 4\pi S\phi'' \tanh(\pi S\phi') + 5\pi^2 S^2\phi'\phi'' \sech^2(\pi  S\phi') \\
& \quad - 2\pi^3 S^3(\phi')^2\phi''\sech^2(\pi S\phi')\tanh(\pi S\phi') + S\frac{\partial \tau_{\text{ext}}^0}{\partial y}.
\end{aligned}
\eeq
Eq.~\eqref{eqn_phi_case2} is a differential equation for $\phi'$, the (non-dimensional) pair density. Its solution describes the pair density distribution in equilibrium when all dipoles form non-local structures as shown in Fig.~\ref{fig_illu_types}(b).

To justify our results for $\phi'$ and $\zeta$ calculated from the continuum model, we also consider the equilibrium states obtained by the discrete dislocation dynamical model. To do that, we simply put $N+1$ pairs of dipoles in the domain $[0,1]$ and let the system evolve to the steady state following Eqs.~\eqref{mobility_law_pos_dimensionless} - \eqref{tau_internal_neg_n_dimensionless}.

For all the simulation results presented in this paper, we lock one dislocation at each end. For example, at the left boundary, we set $p_0 = 0$. There is no strict requirement for $q_0$, except that $q_0\ge0$. Similarly at the right end, we let $q_N = 1$ and $p_N\le1$. By doing this, the total number of dislocation pairs are conserved during the simulation. Correspondingly at the continuum level, this condition is translated by
\beq \label{boundary_lock_con}
\int_0^1 \phi'(t,x) \d x = \phi(t,1) - \phi(t,0) = 1.
\eeq

The temporal derivatives needed for DDD simulations are approximated by using the Euler scheme with time step $\Delta t_{\text{dis}}$ chosen by $\Delta t_{\text{dis}} = 0.025/N$.

Now we compare the results for Equilibrium Type II obtained from the continuum and DDD models. For simplicity, we consider the case when $\partial \tau^0_{\text{ext}} / \partial y$ is a constant.
Thus we integrate Eq.~\eqref{eqn_phi_case2} on both sides to obtain
\beq \label{density_T2_num}
\log\left(\frac{\cosh(\pi \phi'S)}{\phi'}\right) + \left(\frac{\pi\phi'S}{\cosh(\pi\phi'S)}\right)^2 + 3\pi\phi'S\tanh(\pi\phi'S) = C - \pd{\tau^0}{y}\cdot Sx,
\eeq
where $C$ is a constant to be determined by condition \eqref{boundary_lock_con}.

We begin with the case when no stress gradient is applied to the system, i.e. $\partial \tau^0_{\text{ext}} / \partial y = 0$. In this case, Eq.~\eqref{density_T2_num} suggests $\phi' = 1$ and we then obtain $\zeta=1/2$. This means that in the absence of applied stress gradients, all dipoles are uniformly distributed and the dipoles form non-localised structures suggested by the continuum model. To see Equilibrium Type II from the DDD model, one needs $S>0.2465/\phi'$ and $S$ is chosen to be $0.3$ here.

Note that in the DDD model, the pair density is approximated by $\rho_{\text{dis}}((p_n+q_n)/2) = 1/(N(p_{n+1}-p_n))$, and $\zeta$ is approximated by $\zeta_{\text{dis}}((p_n+q_n)/2) = N(q_n-p_n)$.

\begin{figure} [!ht]
\centering
\psfrag{x}{{\small $x$}}
\subfigure[Pair density ]{\includegraphics[width=.4\textwidth]{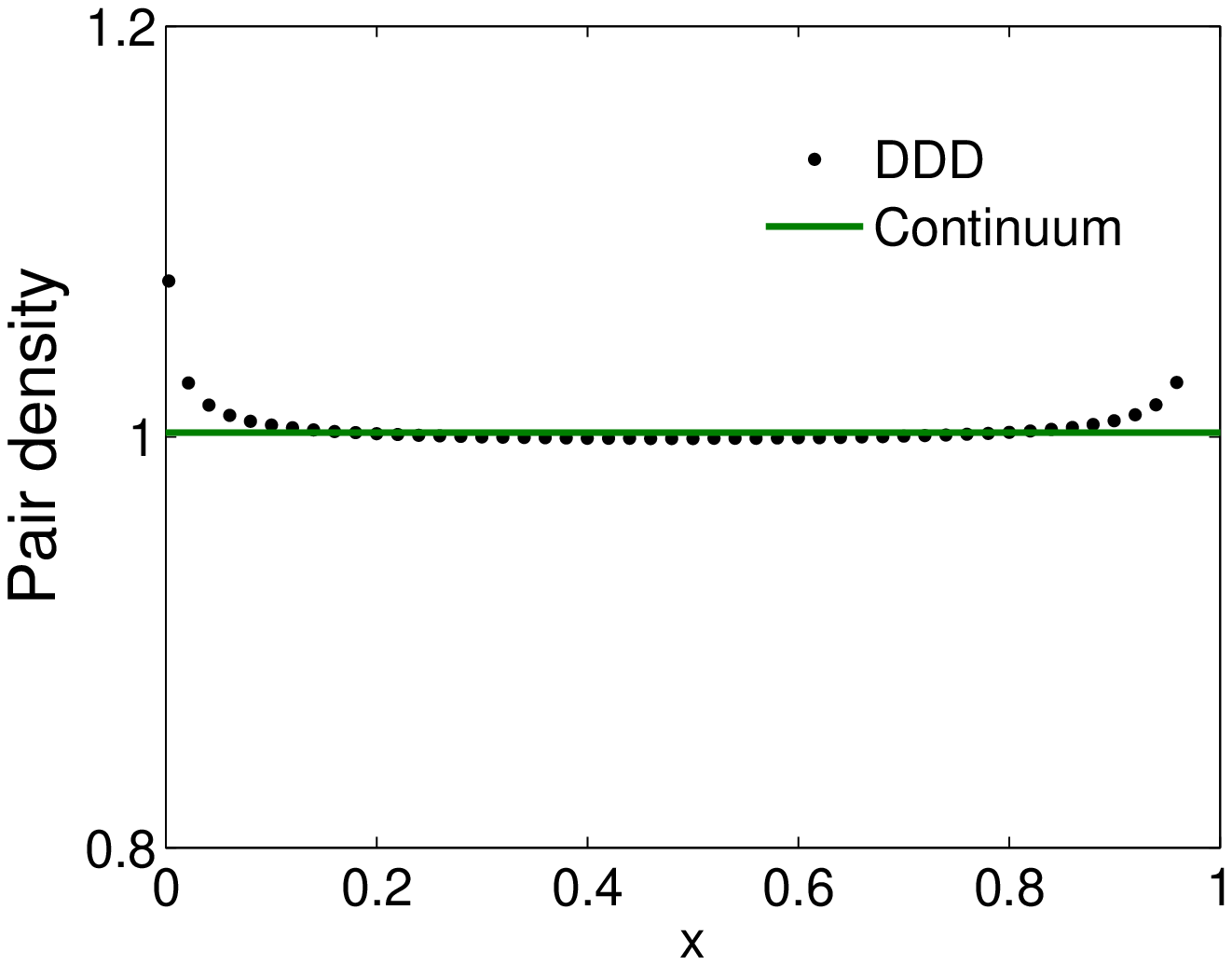}}
\subfigure[Local pattern ]{\includegraphics[width=.4\textwidth]{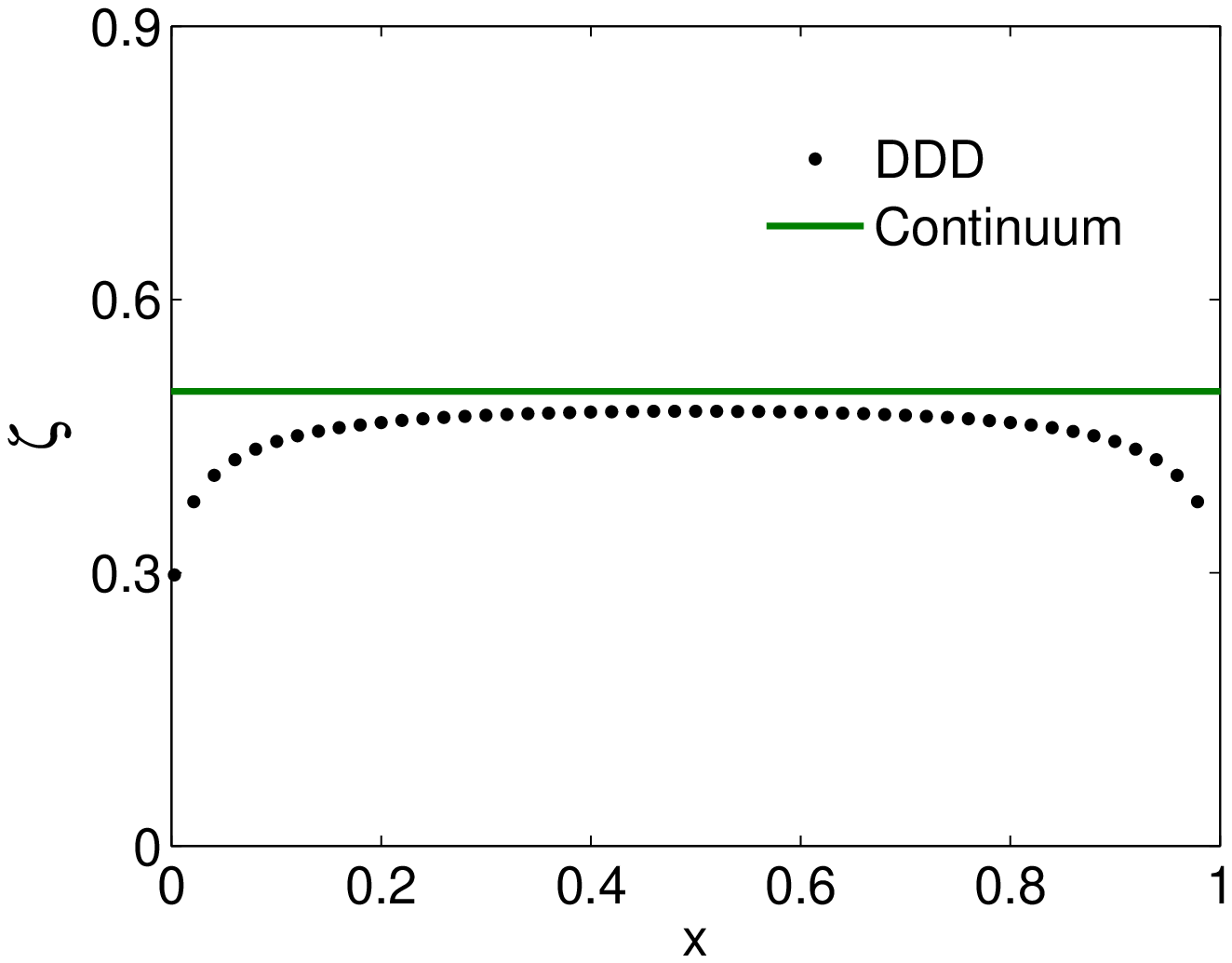}}
\caption{Comparison of the pair density and the pair width Equilibrium Type II with the results from the discrete dislocation dynamical models in the absence of applied stresses or stress gradients. When $S = 0.3$, dipoles form Equilibrium Type II. Here $N = 50$. The dipoles take a uniform distribution within $[0, 1]$. \label{fig_t2_N50_S03_dtaudy0}}
\end{figure}

A comparison of the values of $\phi'$ and $\zeta$ from the discrete and the continuum models is shown in Fig.~\ref{fig_t2_N50_S03_dtaudy0} and good agreement between the two models is seen except near the boundaries.
There is a boundary layer near each end, where the results from the continuum model deviate from its DDD counterpart. This is because the symmetry required for the setting up of the inner region $\Omega_{\text{in}}^n$ given by Eq.~\eqref{inner_region_def} breaks down. However, the goal of this paper is to formulate the collective behaviour of dislocation dipoles in the (relatively vast) interior region. It is suggested by the numerical results shown below that the influence cast by the boundary layers over the accuracy of the continuum approximation in the interior region is limited. Hence the incorporation of boundary layers into the continuum framework will be discussed in future work.

\begin{figure}[!ht]
\centering
\psfrag{x}{{\small $x$}}
\subfigure[Pair density]{\includegraphics[ width=.4\textwidth]{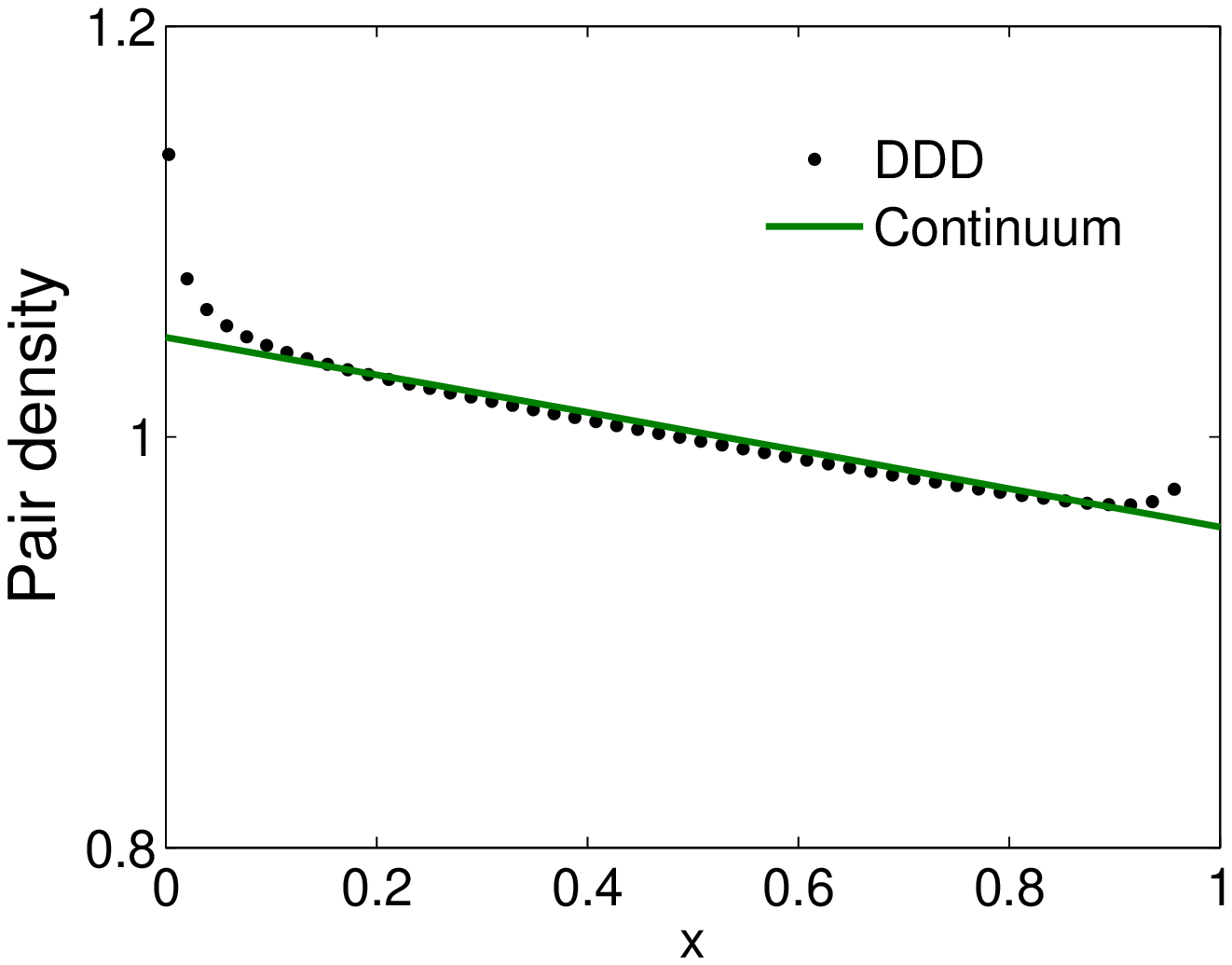}}
\subfigure[Local pattern]{\includegraphics[width=.4\textwidth]{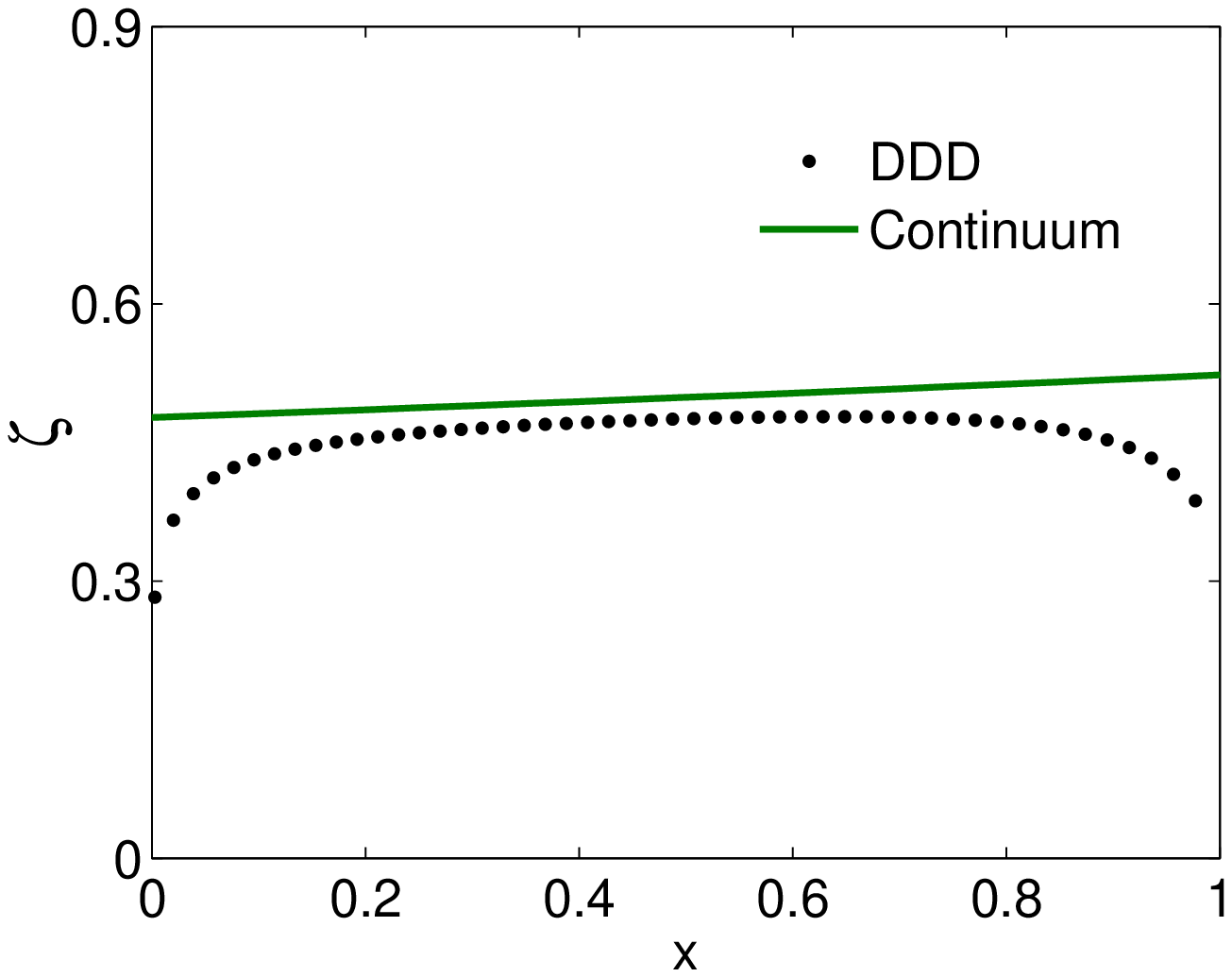}}
\caption{When the system is applied an stress gradient $\partial \tau_{\text{ext}}^0/\partial y=1$, the dipoles of Equilibrium Type II are seen piling-up against the left boundary. Here $S = 0.3$ and $N = 50$. \label{fig_t2_N50_S03_dtaudy1}}
\end{figure}

With a non-vanishing stress-gradient, for example, $\partial \tau_{\text{ext}}^0/\partial y = 1$, one can again calculate $\phi'$ and $\zeta$ with reference to Eq.~\eqref{density_T2_num}. The results from the two models are compared in Fig.~\ref{fig_t2_N50_S03_dtaudy1} and excellent agreement in the interior region is again seen away from the two ends.

\subsubsection{First-order force balance equation for Equilibrium Type III\label{Sec_equilibria_t3}}
Similarly, we study the first-order equation \eqref{eqn_fb_general2}, when all dipoles are in Equilibrium Type III, i.e. condition \eqref{existence_condition_case3} is met. It is recalled that Equilibrium Type III only appears for small $\phi'S$, we consider the asymptotic behaviour of the above equation as $S\rightarrow0$ for simplicity. Thus one can asymptotically solve Eq.~\eqref{equilibrium_type3} to get
\beq \label{zeta_smalls}
\zeta \approx S + \frac{2(\pi\phi')^2S^3}{3}.
\eeq
Eq.~\eqref{zeta_smalls} implies that in this case the pair width is almost the same as the slip plane gap. When these two quantities are identical, we call the resulting dislocation structure a $45^{\circ}$ dipole. In fact, a $45^{\circ}$ dipole is the stable configuration of an isolated pair of dipole. We see from  Eq.~\eqref{zeta_smalls} that when the two slip planes get close to each other (as $S\rightarrow0$), the mutual interaction between the pair partners becomes dominant over the stresses due to all other dislocations, and the dipoles behave as isolated dipolar pairs. Incorporating Eq.~\eqref{zeta_smalls} into the first-order equation \eqref{eqn_fb_general2}, we asymptotically derive an equation for the pair density $\phi'$ in the limit that $S\rightarrow0$ as
\beq \label{eqn_density_smalls}
2\pi^2S^2\phi'\phi'' + S\cdot \frac{\partial \tau_{\text{ext}}^0}{\partial y} = 0.
\eeq
Eqs.~\eqref{zeta_smalls} and \eqref{eqn_density_smalls} are valid only when $S\rightarrow0$. Now we compare their solutions to DDD simulation results to show that they can be used as the governing equations for many dipoles in equilibrium of Type III at the continuum level.

Here we still consider the case when $\partial \tau^0_{\text{ext}}/\partial y$ is constant for simplicity. Hence the pair density distribution $\phi'$ can be solved from Eq.~\eqref{eqn_density_smalls}
\beq \label{density_T3_num}
\phi' = \frac1{\pi S}\sqrt{C-\pd{\tau_{\text{ext}}^0}{y}\cdot Sx},
\eeq
where $C$ is determined by boundary condition \eqref{boundary_lock_con}.

\begin{figure} [!ht]
\centering
\psfrag{x}{{\small $x$}}
\subfigure[Pair density ]{\includegraphics[width=.4\textwidth]{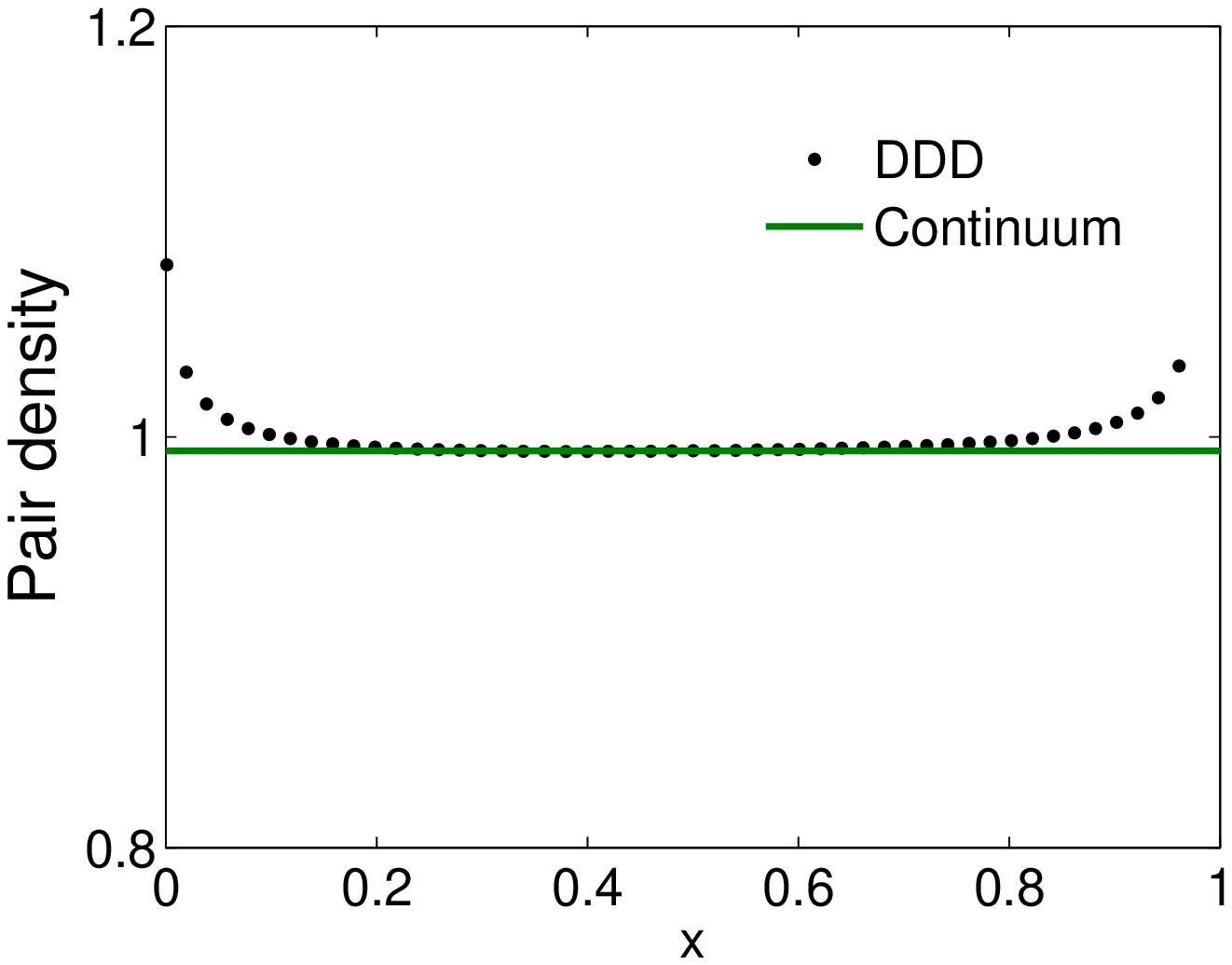}}
\subfigure[Local pattern]{\includegraphics[width=.4\textwidth]{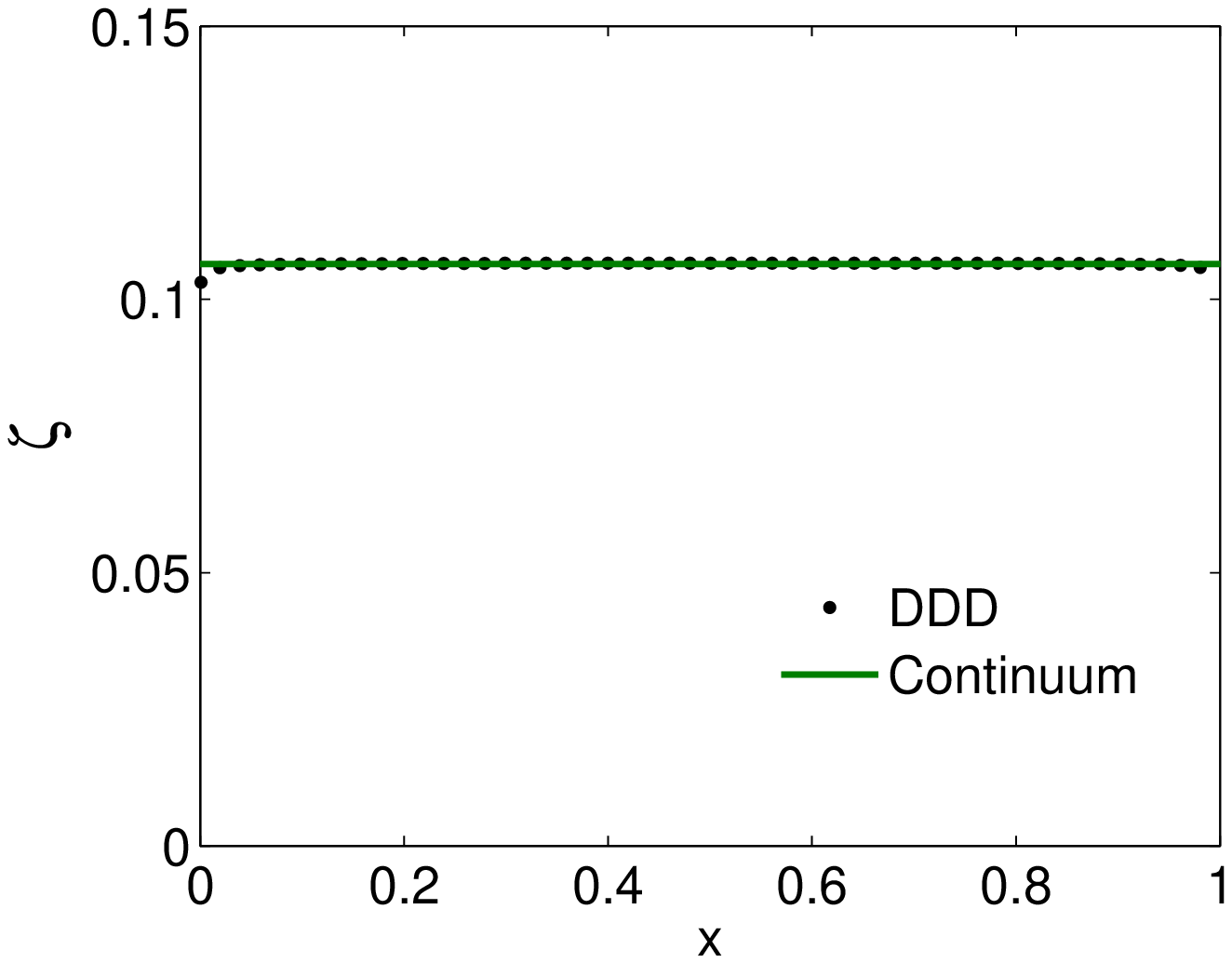}}
\caption{Comparison of results from the continuum and the DDD models for the case where there is no applied stress gradient. When $S = 0.1$, the continuum model suggests that the system takes the equilibrium state of Type III with $\phi'\approx1$ and $\zeta\approx0.1066$. Here $N=50$. \label{fig_t3_N50_S01_dtaudy0}}
\end{figure}

We first investigate the case with no applied stress gradient. From Eq.~\eqref{density_T3_num}, we obtain $\phi'=1$ and $\zeta$ is calculated to be $0.1066$ from Eq.~\eqref{zeta_smalls}. We then compare these results with that from the DDD simulations in Fig.~\ref{fig_t3_N50_S01_dtaudy0}.
Excellent agreement between the two models is seen. Here we find again that in the absence of applied stress gradient, the dipoles are uniformly distributed.

\begin{figure}[!ht]
\centering
\psfrag{x}{{\small $x$}}
\subfigure[Pair density]{\includegraphics[width=.4\textwidth]{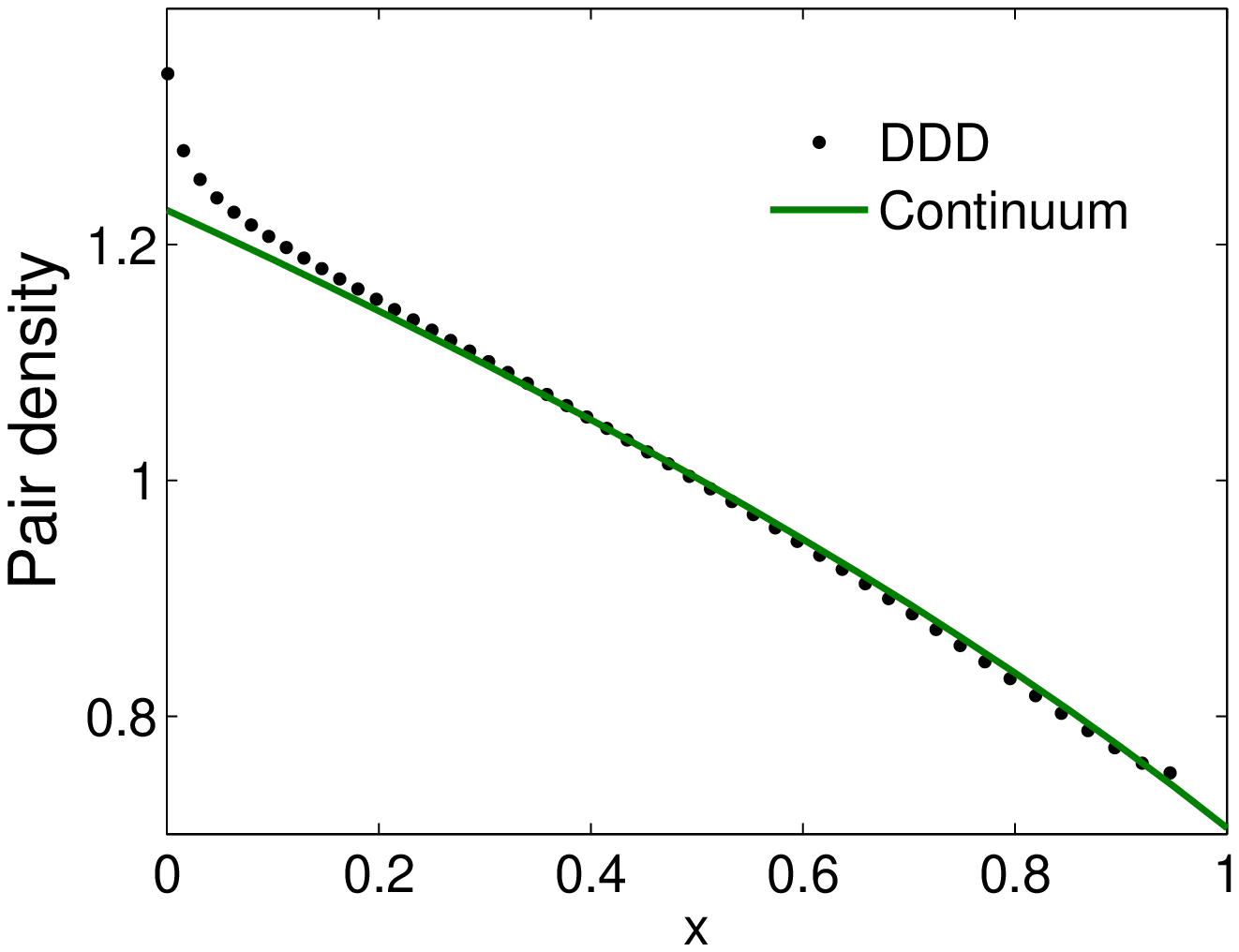}}
\subfigure[Local pattern]{\includegraphics[width=.4\textwidth]{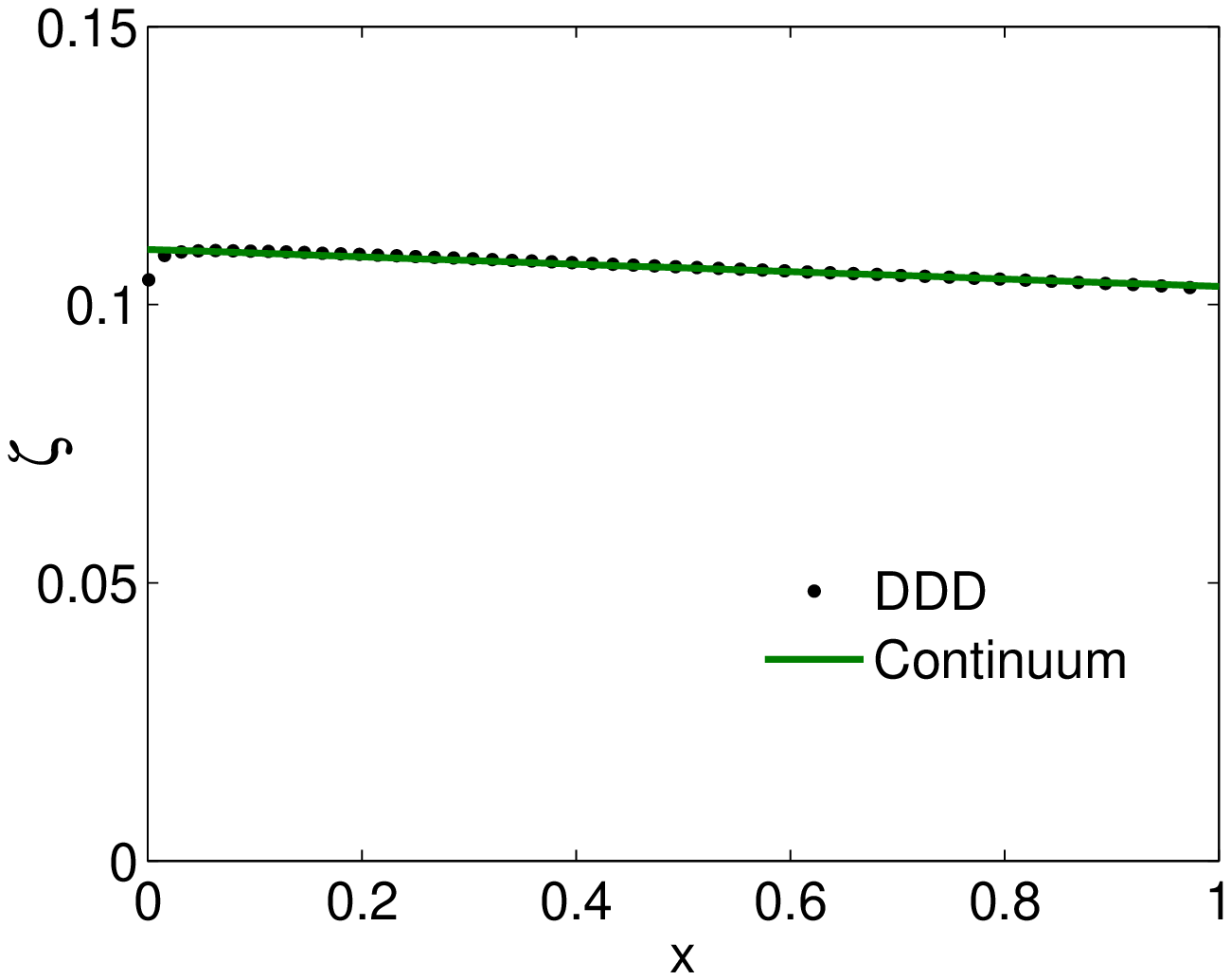}}
\caption{Dipoles of Equilibrium Type III are found piling-up against an applied stress gradient to the left boundary. Here $S = 0.1$, $\partial \tau^0 / \partial y=1$ and $N = 50$. \label{fig_t3_N50_S01_dtaudy1}}
\end{figure}

Now we consider a non-vanishing applied stress gradient set to be $\partial \tau_{\text{ext}}^0/\partial y = 1$. By using Eqs.~\eqref{zeta_smalls} and \eqref{eqn_density_smalls}, we plot $\phi'$ and $\zeta$ against $x$ in Fig.~\ref{fig_t3_N50_S01_dtaudy1} and they are shown agreeing well with the outcomes from the underlying DDD model.
The comparison results shown above suggest that we can use Eqs.~\eqref{zeta_smalls} and \eqref{eqn_density_smalls} to describe the collective behaviour of a row of dislocation dipoles in equilibrium of Type III.

\subsubsection{Equilibria of mixed types}
\begin{figure}[!ht]
\centering
\psfrag{x}{{\small $x$}}
\subfigure[Pair density]{\includegraphics[width=.45\textwidth]{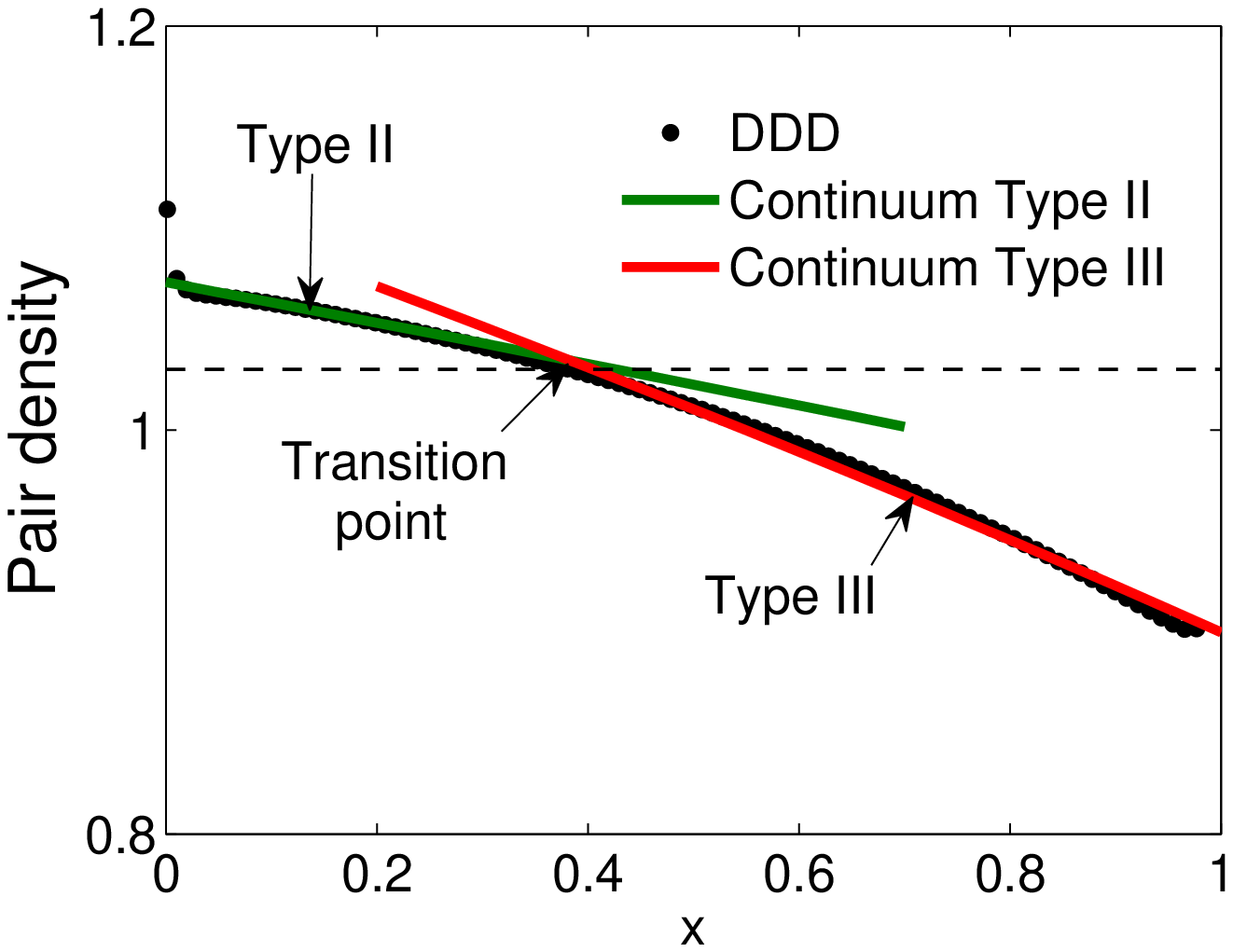}}
\subfigure[Local pattern]{\includegraphics[width=.45\textwidth]{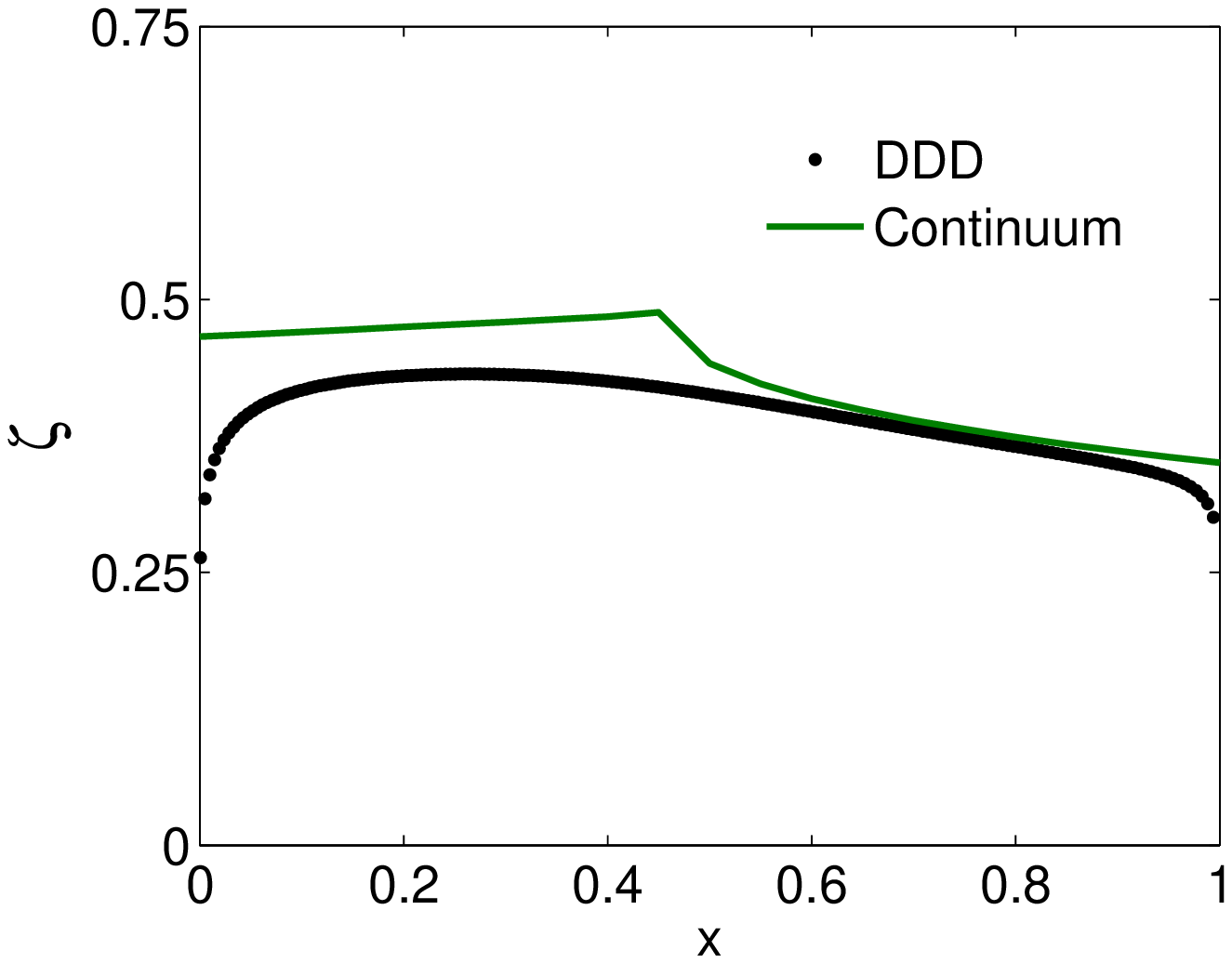}}
\caption{When $S=0.24$, Equilibrium Type II and III are found co-exist. Near the left boundary, the dipoles take the equilibrium of Type II. A natural transition from Type II to III is seen roughly where the pair density drops below the dashed line characterised by $\phi'\approx1.03$. Here $N = 100$. \label{fig_mix_N50_S024_dtaudy1}}
\end{figure}

According to Eq.~\eqref{existence_condition_case3}, $\phi'S=0.2465$ characterises the transition between Equilibrium Type II and III. Therefore, when the value of $\phi'-0.2465/S$ changes its sign, there should be a change in equilibrium patterns as suggested by the continuum model. This is actually observed in Fig.~\ref{fig_mix_N50_S024_dtaudy1}, where $S$ is set to be $0.24$ and $N$ is chosen to be $100$. It is seen from Fig.~\ref{fig_mix_N50_S024_dtaudy1} that the dipoles take Equilibrium Type II near the left boundary, and a transition from Type II to III is found taking place away from the left end. The continuum model suggests that the transition should happen when $\phi' = 0.2465/S \approx 1.03$, which gives rise to the dashed line in Fig.~\ref{fig_mix_N50_S024_dtaudy1}. It can be checked that Equilibrium Type III roughly emerges where $\phi'$ drops below the dashed line. In Fig.~\ref{fig_mix_N50_S024_dtaudy1}, it can also be seen that the values of the pair density agree well for both equilibrium types, while there is roughly a $10\%$ variance in $\zeta$ for Equilibrium Type II with the change of equilibrium type not so easily determined. We will see later that increasing $N$ will bring down the deviation in $\phi'$ and $\zeta$ between the continuum and the DDD models.

\subsubsection{Summary}
To summarise, a row of dipoles may form two types of stable equilibria if the applied stress vanishes on $y=0$. When $\phi'S\ge0.2465$, the resulting equations at the continuum level of the pair density $\phi'$ and (rescaled) pair width $\zeta$ are derived to be Eq.~\eqref{eqn_phi_case2} and $\zeta=1/(2\phi')$. When $0 < \phi'S < 0.2465$, the collective behaviour of a row of dipoles can be approximately described by Eqs.~\eqref{zeta_smalls} and \eqref{eqn_density_smalls}.

\subsection{Equilibria under arbitrary externally-applied stresses\label{Sec_eqn_tauext}}
Now we generalise our discussion to the case where the leading order of the external resolved shear stress is non-vanishing, i.e. $\tau^0_{\text{ext}} \sim \CO(1)$. In this case, Eq.~\eqref{eqn_fb_general1} may not be solved explicitly. However, some analysis can still be done to understand the resulting equilibrium configurations.

\begin{figure}[!ht]
  \centering
  \includegraphics[width=.55\textwidth]{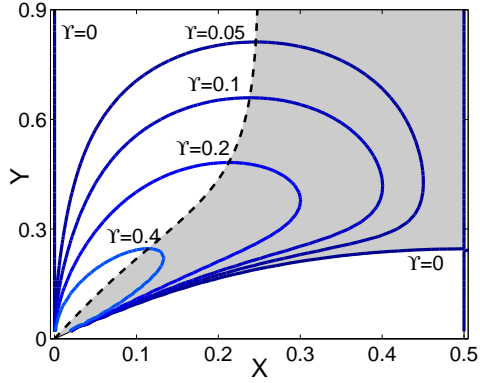}
  \caption{Given any $\Upsilon = \tau^0_{\text{ext}}/(\pi\phi')$, a pair of $(X,Y)$, which satisfies Eq.~\eqref{eqn_zeta_ext_nonzero} should sit on the contour $-G_0(2\pi\phi'\zeta,2\pi\phi'S)$ with height $\Upsilon$. For each $\Upsilon$, there exists a $Y^*$ (attained at $X^*$ say) such that $Y \le Y^*$. The locus of such $(X^*,Y^*)$ lies on the dashed curve. For any $Y<Y^*$ (under a given $\Upsilon$), there are two possible values for $X$. Only those $(X,Y)$ falling in the shaded region correspond to stable configurations. \label{fig_contour_tau_ext}}
\end{figure}

If we use the expression for $G_0$ defined by Eq.~\eqref{term_G0}, we rewrite Eq.~\eqref{eqn_fb_general1} as
\beq \label{eqn_zeta_ext_nonzero}
G_0(2\pi\phi'\zeta,2\pi\phi'S) + \frac{\tau^0_{\text{ext}} }{\pi\phi'} = 0.
\eeq
Eq.~\eqref{eqn_zeta_ext_nonzero} describes the inter-relation of three quantities, $\zeta\phi'$, $S\phi'$ and $\tau^0_{\text{ext}}/\phi'$ and we define $
X = \zeta\phi'$, $Y = S\phi'$, and $\Upsilon = \tau^0_{\text{ext}}/(\pi\phi')$
to facilitate further analysis. As discussed in \S\ref{Sec_fb_leading_order_zero}, $X$ and $Y$ measure respectively the pair width and the slip plane gap, both scaled by the spacing between the neighbouring dipolar centers. Thus Eq.~\eqref{eqn_zeta_ext_nonzero} can be written as $- G_0(2\pi X,2\pi Y) = \Upsilon$, which suggests that the inter-relation between $X$ and $Y$ for a given $\Upsilon$ can be visualised by the contours of $-G_0(2\pi X,2\pi Y)$ as shown in Fig.~\ref{fig_contour_tau_ext}.
It can be observed that on each contour, there exists a $Y^*$ (attained at $X^*$ say) such that $Y \le Y^*$, and the locus of such $(X^*,Y^*)$ sits on the dashed curve in Fig.~\ref{fig_contour_tau_ext}. This means the solution $\zeta$ to Eq.~\eqref{eqn_zeta_ext_nonzero} conditionally exists. Given $Y=S\phi'$, the solution for $X=\zeta\phi'$ satisfying Eq.~\eqref{eqn_zeta_ext_nonzero} exists for
\beq \label{critical_stress_external}
\tau_{\text{ext}}^0 = \pi\phi'\Upsilon \le \pi\phi'\cdot|G_0(2\pi X^*,2\pi Y^*)|.
\eeq
The physical interpretation of Eq.~\eqref{critical_stress_external} is that a dipole breaks down to two monopoles when the external stress is large.

It is also observed from Fig.~\ref{fig_contour_tau_ext} that there exist two choices for $X$ when $Y<Y^*$. One way to identify the stability of the candidate solutions is by investigating the local minima of the generalised free energy density $\CF$ by Eq.~\eqref{free_energy} with respect to $\zeta$. Here we find that the larger one gives rise to a stable equilibrium state after checking with the numerical results to be shown later. Hence we conclude that only those $(X,Y)$ falling into the shaded region in Fig.~\ref{fig_contour_tau_ext} correspond to stable configurations.

It is worth noting that $\Upsilon>0$ is considered in the analysis presented above. When $\Upsilon<0$, we simply let $X<0$ and same conclusion will be drawn.

The above analysis provides us some insight to the equilibrium configurations under an arbitrary externally-applied stress. Nevertheless, to find $\zeta$ and $\phi'$ satisfying Eq.~\eqref{eqn_fb_general1} and \eqref{eqn_fb_general2}, one has to turn to numerical methods.

\section{Comparison of the continuum model with its underlying DDD model\label{Sec_continuum_DDD_comp}}
Now we compare the simulation results obtained by applying the continuum model and the DDD model to same dynamical processes. For simulations at the discrete level, the set-up and procedure is as in \S\ref{Sec_type2}. To numerically implement the continuum model, we discretise Eqs.~\eqref{eqn_zeta_static} and \eqref{eqn_phi_evo_ts} with step $\Delta x$ in space and $\Delta t_{\text{con}}$ in time. At each time step, we use the following procedure to update the two variables $\phi$ and $\zeta$. With $\phi$ computed from the previous step, we (numerically) solve Eq.~\eqref{eqn_zeta_static} to update the value for $\zeta$ at each spatial grid point. It is worth noting that following the analysis in \S\ref{Sec_eqn_tauext}, we need to ensure the computed $\zeta$ is associated with a stable equilibrium state. Then we use Eq.~\eqref{eqn_phi_evo_ts} to update $\phi$. For the simulation results presented here, $\Delta t_{\text{con}}$ was chosen to be $1.25\Delta x^2$.

Our goal here is to check the accuracy and the efficiency of the continuum model with reference to its underlying DDD model. To measure accuracy, we define
\beq\label{diff_rel_phi}
\text{Err}_{\phi'} = \max_{x\in I}\frac{\phi'-\rho_{\text{dis}}}{\rho_{\text{dis}}},
\eeq
where $\rho_{\text{dis}}$ denotes the density computed by the DDD simulations; we choose $I=[0.1,0.9]$ to avoid the inherent difference between the two methods near the two boundaries. Thus $\text{Err}_{\phi'}$ is used as a measurement of the relative error of the pair density caused by the discrete-to-continuum transition. In a similar sense, we define a measurement of the relative error of the pair width by
\beq\label{diff_rel_zeta}
\text{Err}_{\zeta} = \max_{x\in I}\frac{\zeta-\zeta_{\text{dis}}}{\zeta_{\text{dis}}}.
\eeq

\begin{figure}[!ht]
  \centering
  \psfrag{x}{{\small $x$}}
  \subfigure[$t=0$]{\includegraphics[width=.32\textwidth]{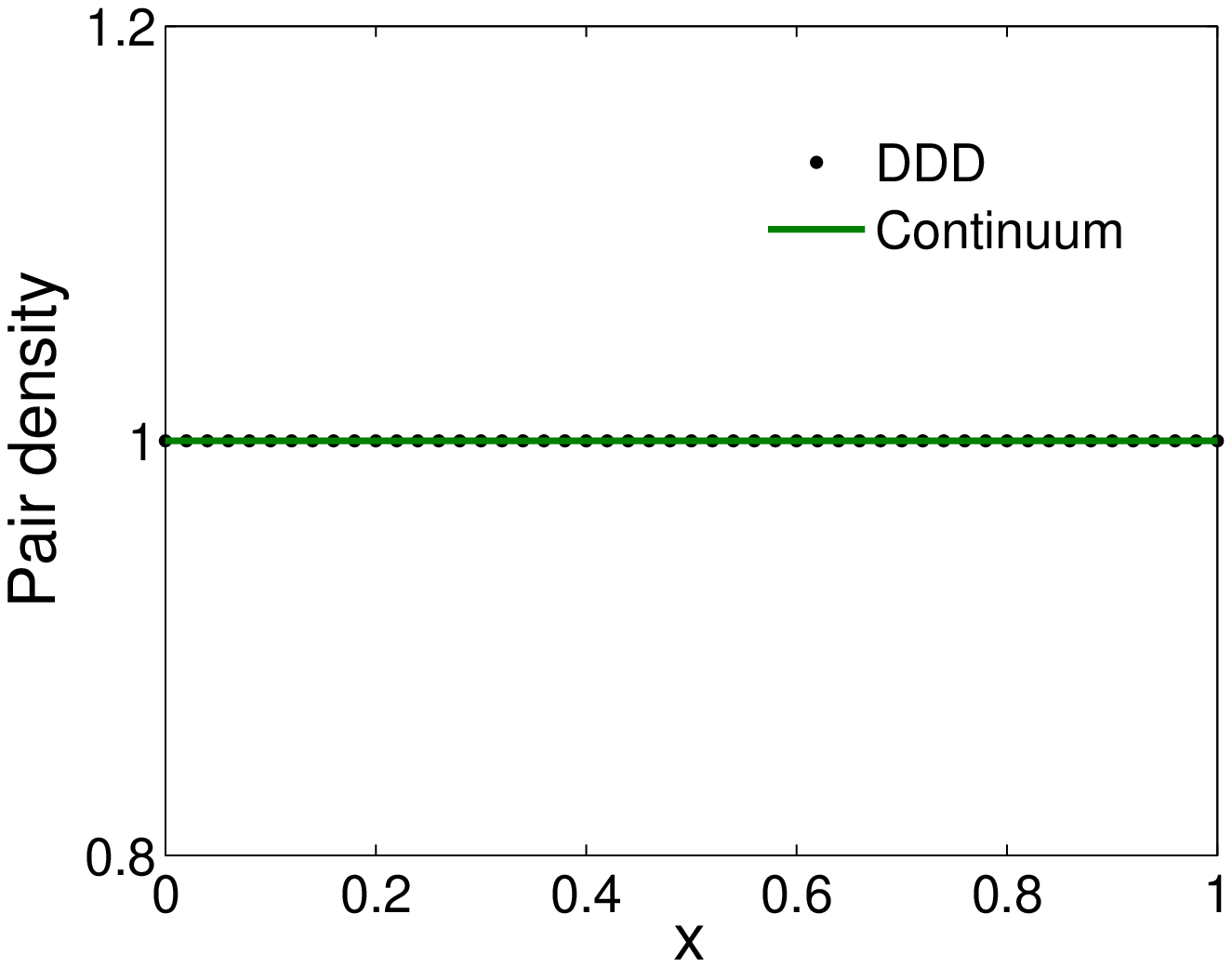}}
  \subfigure[$t=1$]{\includegraphics[width=.32\textwidth]{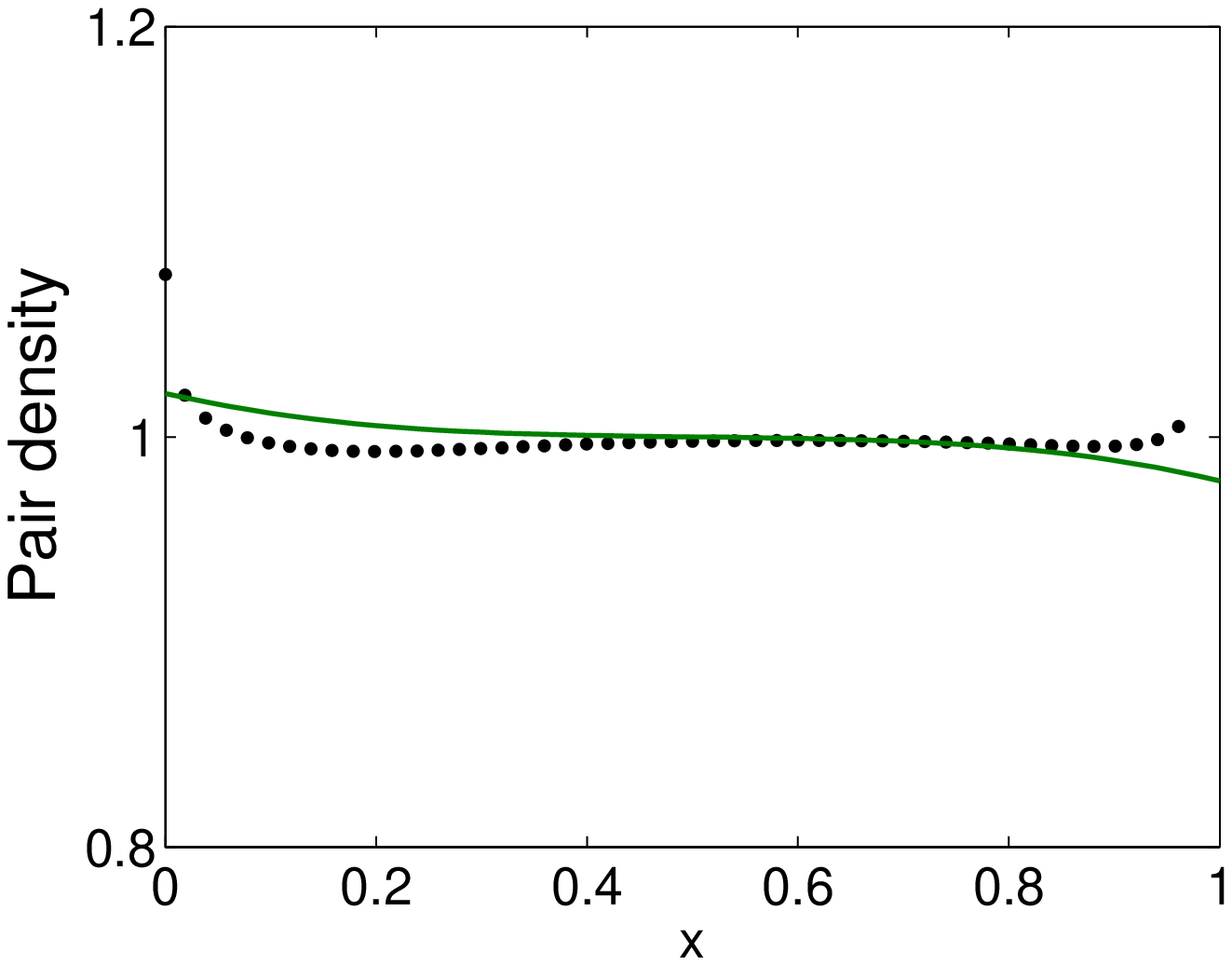}}
  \subfigure[$t=2$]{\includegraphics[width=.32\textwidth]{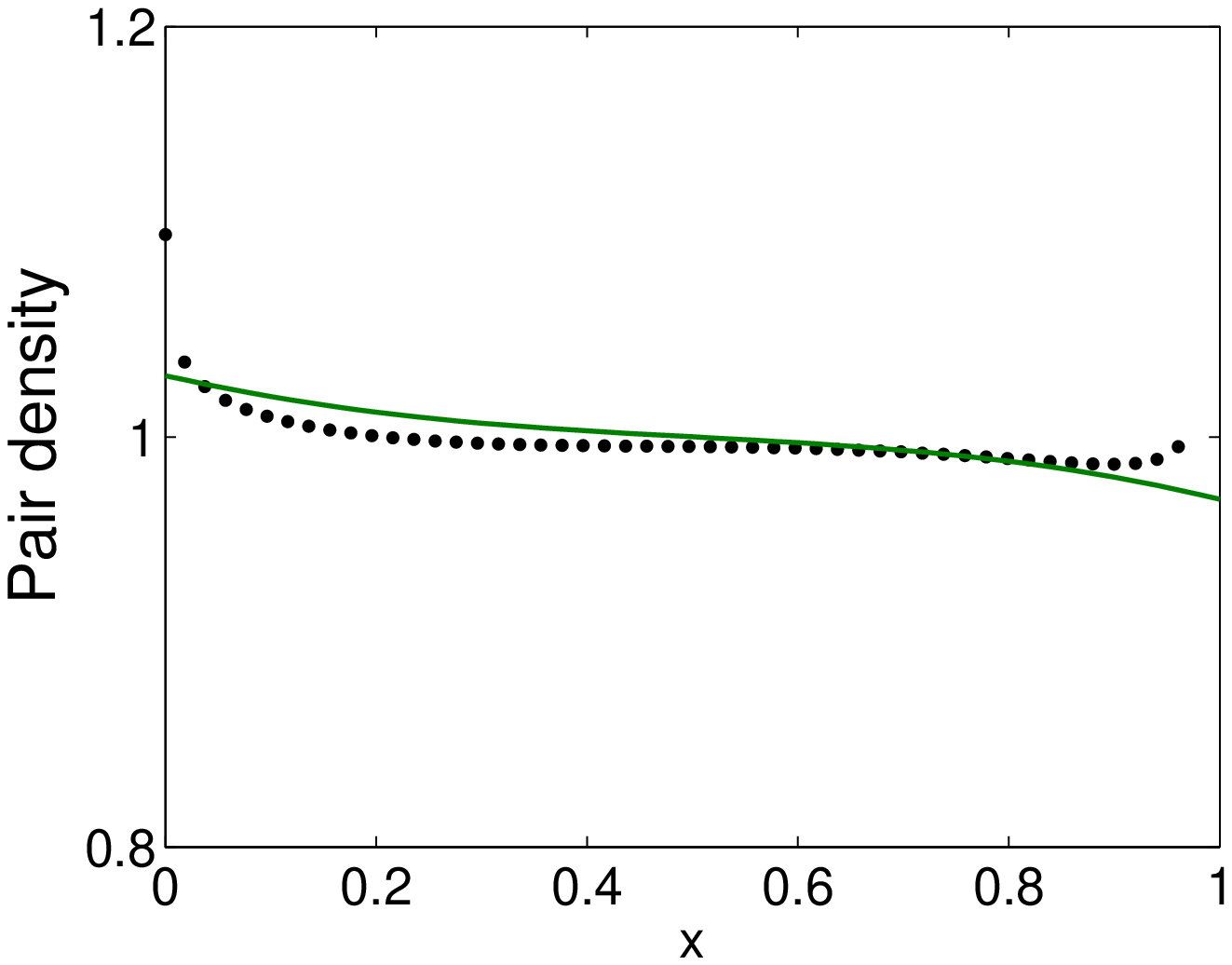}}
  \subfigure[$t=5$]{\includegraphics[width=.32\textwidth]{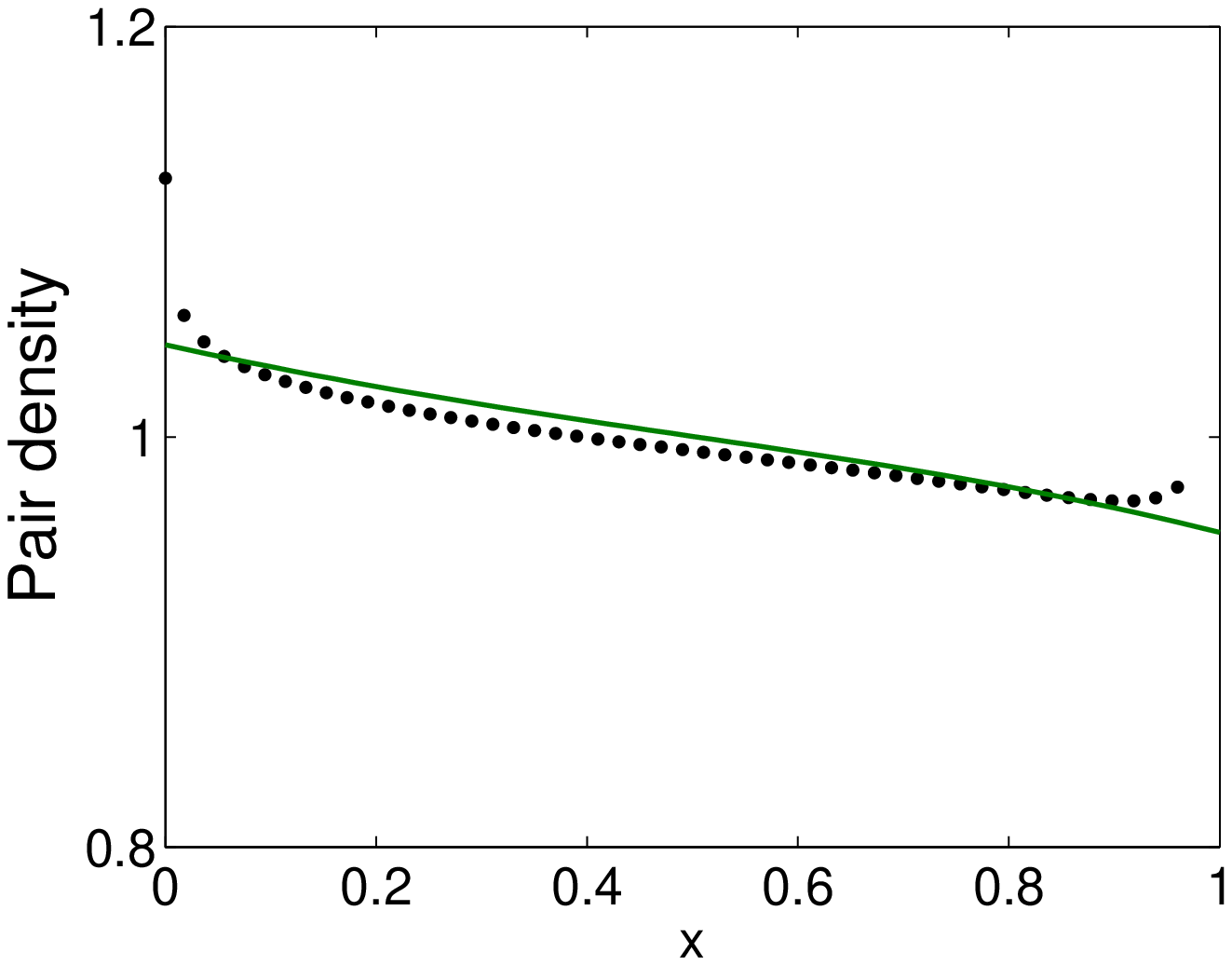}}
  \subfigure[$t=10$]{\includegraphics[width=.32\textwidth]{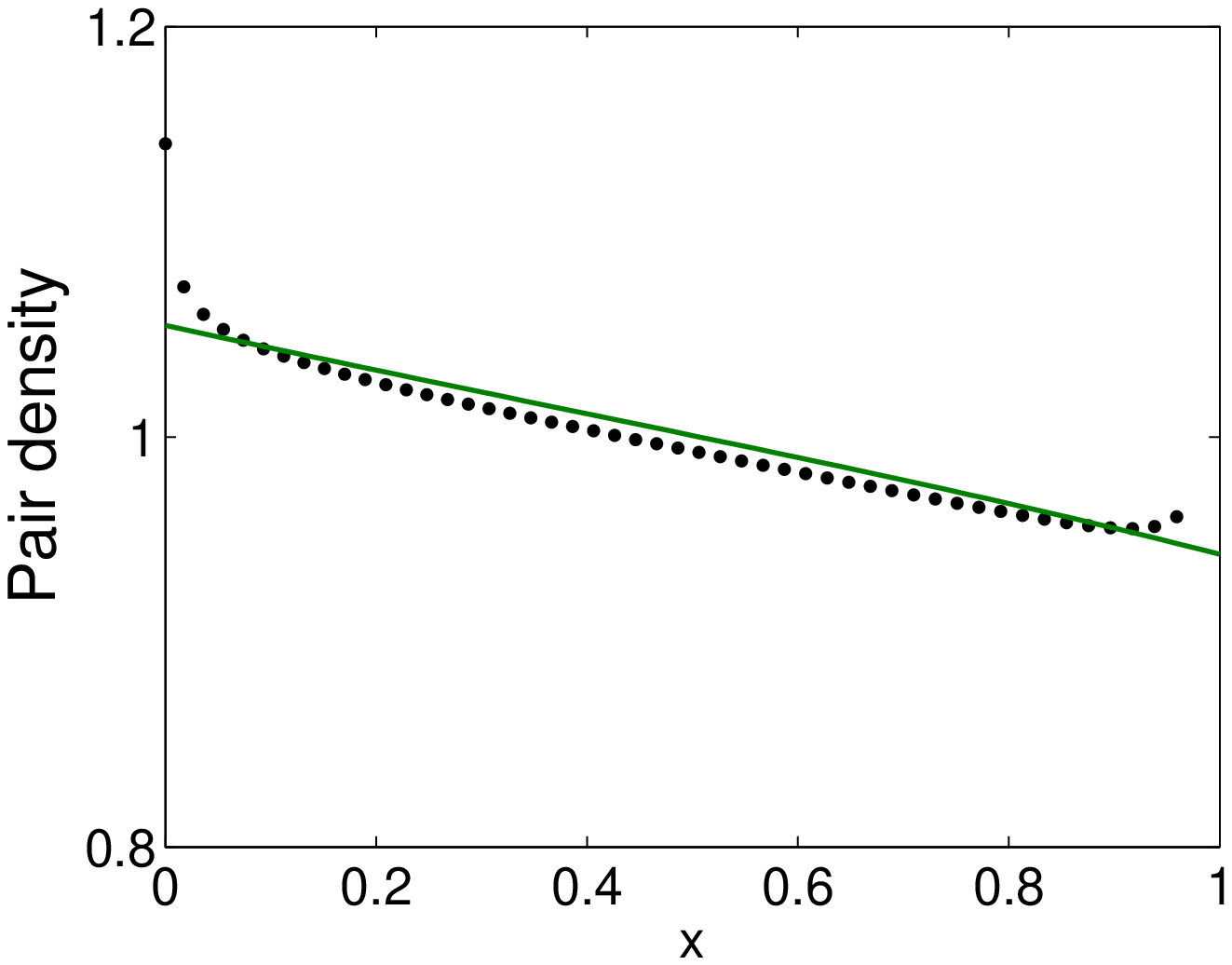}}
  \subfigure[$t=26.4$]{\includegraphics[width=.32\textwidth]{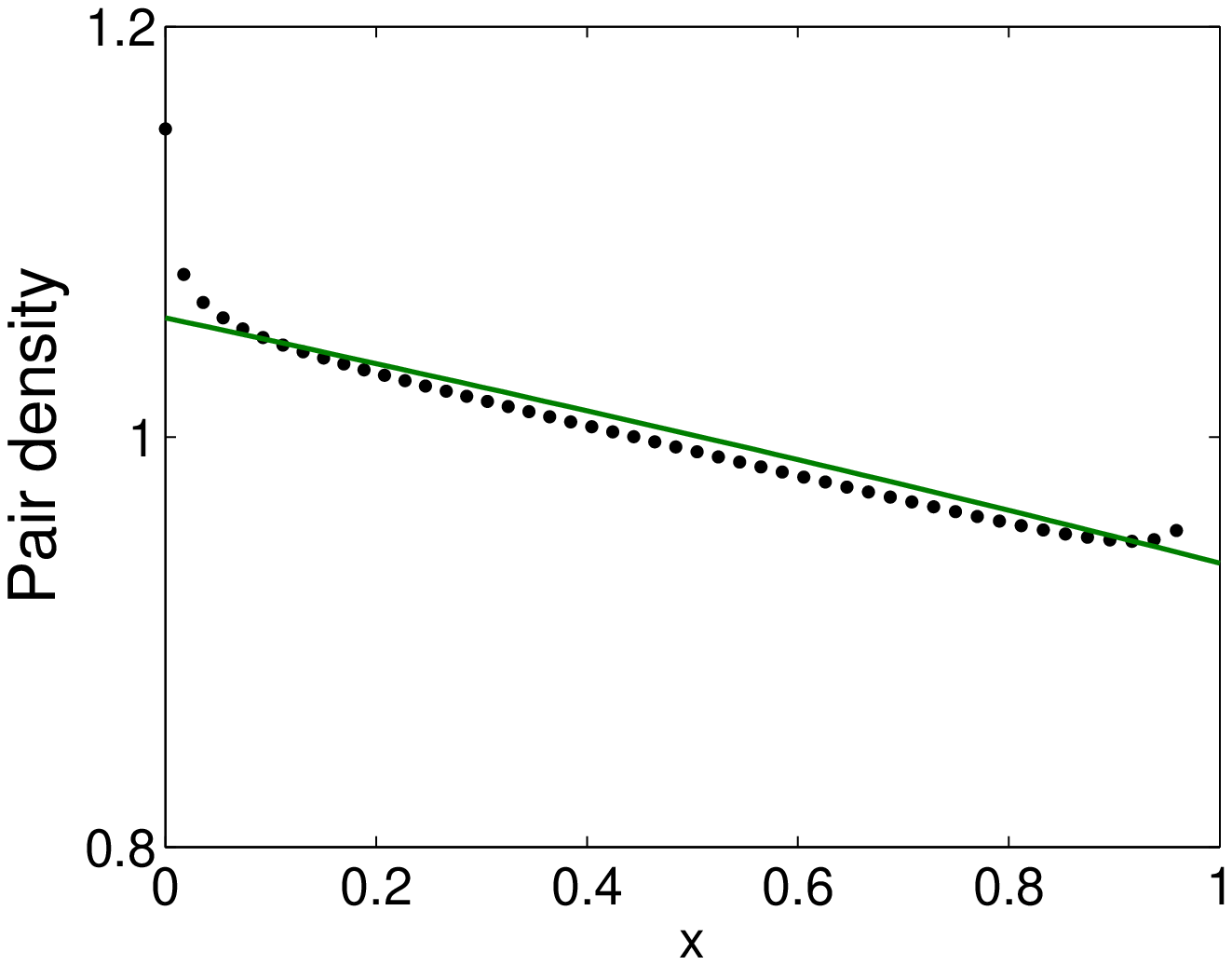}}
  \caption{Snap shots of the pair density obtained from the DDD and the continuum methods at $t=0$, $1$, $2$, $5$, $10$ and $26.4$, where $t$ is measured in unit $2\pi(1-\nu)L^2/(m_{\text{g}}\mu b^2)$.\label{fig_com_S03}}
\end{figure}

The parameters chosen for the first set of numerical examples are $S=0.3$, $N=50$, $\tau^0_{\text{ext}}=0.5$ and $\partial \tau^0_{\text{ext}}/\partial y=1$. In Table.~\ref{table_compare_dis_con_S03}, $\text{Err}_{\phi'}$ and $\text{Err}_{\zeta}$ at various times are listed.
\begin{table}[!ht]
  \centering
  \begin{tabular}{c|cccccc}
  $t$ & 1 & 2 & 5 & 10 & 20 & 26.4\\
  \hline
  $\text{Err}_{\phi'}$ & 0.0150 & 0.0117 & 0.0088 & 0.0077 & 0.0079 & 0.0079\\
  $\text{Err}_{\zeta}$ & 0.0797 & 0.0801 & 0.0810 & 0.0815 & 0.0818 & 0.0818
  \end{tabular}
  \caption{Defined by Eq.~\eqref{diff_rel_phi}, $\text{Err}_{\phi'}$ provides a measurement of the relative error of the pair density caused by the discrete-to-continuum transition. Similarly $\text{Err}_{\zeta}$ given by Eq.~\eqref{diff_rel_zeta} provides a measurement of the relative error of the pair width $\zeta/N$. Here $S=0.3$, $\tau^0_{\text{ext}}= 0.5$, $\partial \tau^0_{\text{ext}}/\partial y = 1$ and $N=50$. Here $t$ is measured in unit $2\pi(1-\nu)L^2/(m_{\text{g}}\mu b^2)$. Simulations by the two models both stop at $t=26.4$, when the difference in the dislocation positions in DDD simulations between this and the previous time step is no more than $10^{-5}\Delta t_{\text{dis}}$. $\text{Err}_{\phi'}$ and $\text{Err}_{\zeta}$ are listed at various times. \label{table_compare_dis_con_S03}}
\end{table}
Note that the time $t$ in Table~\ref{table_compare_dis_con_S03} is measured in unit $2\pi(1-\nu)L^2/(m_{\text{g}}\mu b^2)$ with $L$ recalled to be the computational domain size. The simulations based on both the continuum and DDD models are stopped at $t=26.4$, when the difference in the dislocation positions in DDD simulations between this and the previous time step is no more than $10^{-5}\Delta t_{\text{dis}}$. We see that the relative error in the pair density at different stages is no more than $1.5\%$, while the relative error in pair width is roughly $8\%$. In Fig.~\ref{fig_com_S03}, snap shots of pair density by using the two methods at $t=0$, $1$, $2$, $5$, $10$ and $26.4$ are shown.

We also check the efficiency of the continuum model by keeping all other parameters unchanged while increasing the total number of dislocations $N$. For this purpose, we introduce two quantities $T_{\text{con}}$ and $T_{\text{dis}}$, which denote the wall-clock time it takes a simulation to reach the steady state by using the continuum and DDD models, respectively. Thus $T_{\text{con}}/T_{\text{dis}}$ becomes a measurement of the computational efficiency of using the continuum model against its underlying DDD model. The smaller this value is, the higher efficiency the continuum model displays.

\begin{figure}[!ht]
  \centering
  \subfigure[]{\includegraphics[width=.45\textwidth]{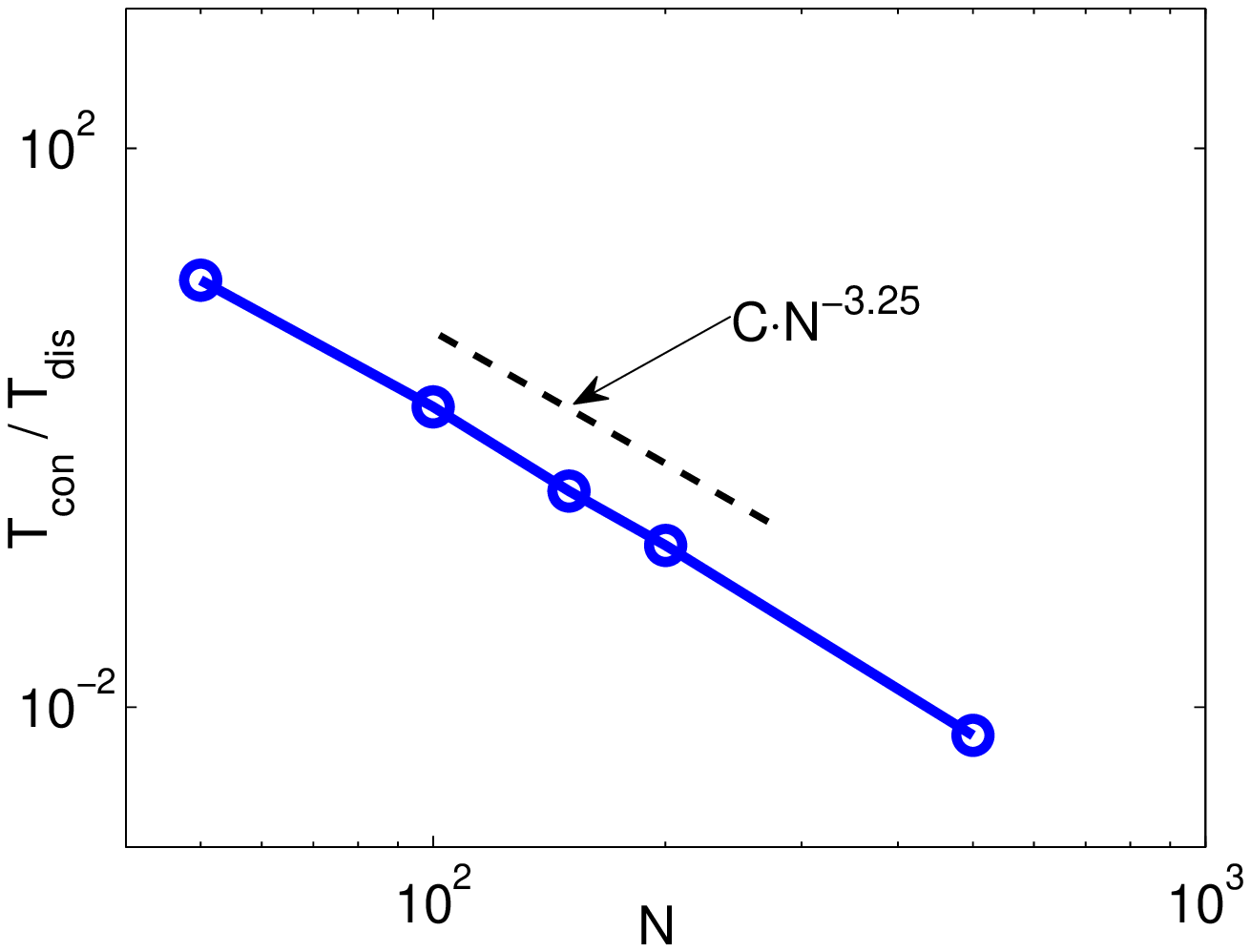}}
  \subfigure[]{\includegraphics[width=.45\textwidth]{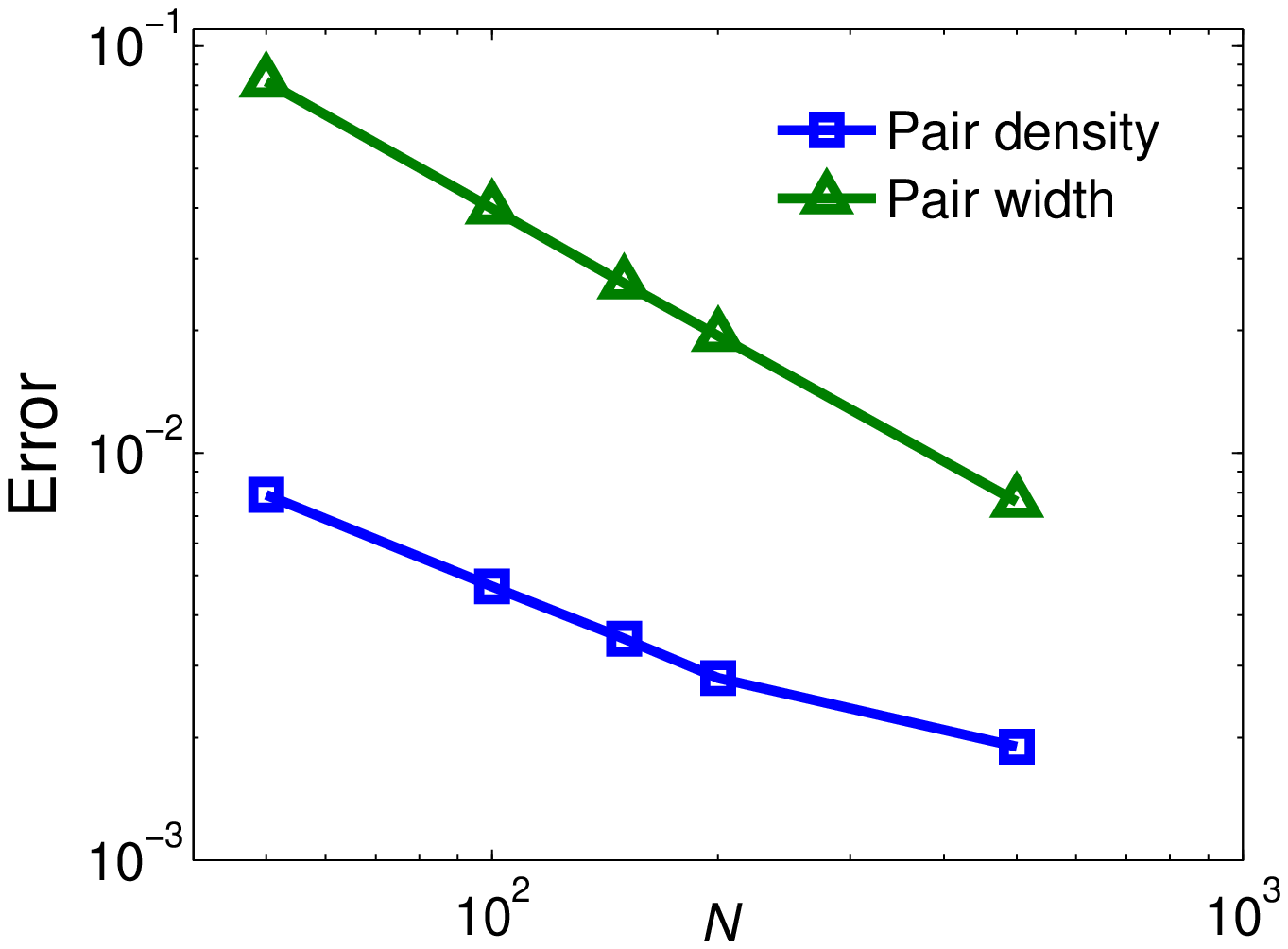}}
  \caption{(a) $T_{\text{con}}/T_{\text{dis}}$ provides a measurement to the computational efficiency exhibited by the continuum model compared to its underlying DDD model. The smaller this value is, the more efficient the continuum model is. (b) The upscaling errors of the pair density $\phi'$ and the pair width defined by Eqs.~\eqref{diff_rel_phi} and \eqref{diff_rel_zeta}, respectively, as the systems attain their steady states with various $N$. \label{fig_compare_N}}
\end{figure}

The comparison between the two models for different $N$ is shown in Fig.~\ref{fig_compare_N}.
In Fig.~\ref{fig_compare_N}(a), $T_{\text{con}}/T_{\text{dis}}$ is found scaling with $N$ at an exponent of roughly $-3.25$. When the total number of dislocation pairs is increased to $500$, the time it takes for the continuum model to reach the steady state is roughly $1\%$ of that needed for the DDD model. This suggests that the continuum model becomes extremely efficient for a large $N$ compared to its DDD counterpart. The greater efficiency displayed by the continuum model can be attributed to the fact that an increase in $N$ only brings up the computational intensity of performing the DDD simulations, as the governing equations \eqref{eqn_zeta_static} and \eqref{eqn_phi_evo_ts} for the continuum model are both independent of $N$.

To check the accuracy of the continuum model, we also plot $\text{Err}_{\phi'}$ and $\text{Err}_{\zeta}$ given by Eqs.~\eqref{diff_rel_phi} and \eqref{diff_rel_zeta}, respectively against $N$ in Fig.~\ref{fig_compare_N}(b). The coarse-graining errors (in the interior region) for both quantities drop with an increasing $N$. When $N$ is $500$, the coarse graining error of $\zeta$ in the interior region measured by Eq.~\eqref{diff_rel_zeta} becomes as good as no more than $1\%$. This is sensible since the continuum model is obtained by taking the asymptotic limit as $N\rightarrow\infty$. An increased $N$ effectively brings down the truncation errors.

When the rescaled slip plane gap $S$ is small, the simulation can be speeded up using the asymptotic solutions to Eq.~\eqref{eqn_zeta_static}, rather than numerically solving Eq.~\eqref{eqn_zeta_static} at each time step. In this scenario, the governing equations at the continuum level can be asymptotically simplified to
\begin{subequations}
\beq \label{zeta_smalls_tauext}
\zeta = S - 2S^2\tau_{\text{ext}}^0 + \left(2(\tau_{\text{ext}}^0)^2 + \frac{2(\pi\phi')^2}{3}\right)S^3
\eeq
and
\beq \label{phi_smalls_tauext}
\pd{\phi}{t_{\text{s}}} - \left(\pi^2S^2\phi''\phi'+ \frac{\zeta}{2}\cdot \pd{\tau^0_{\text{ext}}}{x} + \frac{S}{2} \cdot \pd{\tau^0_{\text{ext}}}{y}\cdot\frac{\partial \phi}{\partial x}\right) \pd{\phi}{x} = 0.
\eeq
\end{subequations}

In Table~\ref{table_compare_dis_con_S01}, the coarse-graining errors for the pair density and the pair width are shown with $N=50$ and $S=0.1$.
\begin{table}[!ht]
  \centering
  \begin{tabular}{c|cccccccc}
  $t$ & 5 & 10 & 20 & 50 & 100 & 200 & 250 & 300\\
  \hline
  $\text{Err}_{\phi'}$ & 0.0060 & 0.0064 & 0.0068 & 0.0074 & 0.0130 & 0.0208 & 0.0221 & 0.0227\\
  $\text{Err}_{\zeta}$ & 0.0179 & 0.0181 & 0.0184 & 0.0189 & 0.0188 & 0.0185  & 0.0184 & 0.0184
  \end{tabular}
  \caption{The coarse-graining errors of the pair density distribution and the pair width at various time slots. Here $S=0.1$, $\tau^0 = 0.5$, $\partial \tau^0/\partial y=1$, $N=50$. Here $t$ is measured in unit $2\pi(1-\nu)L^2/(m_{\text{g}}\mu b^2)$. \label{table_compare_dis_con_S01}}
\end{table}
The upscaling errors are found well controlled during the simulations.

\section{Conclusion and further discussion\label{Sec_conclusion}}
\subsection{Conclusion}
In this paper, we have studied the collective behaviour of a row of dislocation dipoles using matched asymptotic analysis. The discrete-to-continuum transition is facilitated by the introduction of two field variables, the dislocation pair density potential $\phi$ and the dislocation pair width $\zeta$. The equilibrium state at the continuum level is governed by Eqs.~\eqref{eqn_fb_general1} and \eqref{eqn_fb_general2}, while the dynamics at the continuum level is given by Eqs~\eqref{eqn_zeta_static} and \eqref{eqn_phi_evo_ts}. The following conclusions are drawn based on the analysis and the numerical implementation to the continuum model.

Dislocation dipoles are found roughly uniformly distributed in the absence of applied stress gradients, and to pile up against a lock when a stress gradient is applied.

When the externally applied stress is zero on the primary slip plane $y=0$, we found three possible equilibrium patterns (as shown in Fig.~\ref{fig_illu_types}), whose stability depends on the value of $\phi'S$, the ratio of the slip plane gap to the pair center spacing. If $\phi'S$ is big (condition \eqref{existence_condition_case3} breaks down), non-localised structures (Equilibrium Type II) are the stable configurations. When $\phi'S$ falls below the critical value 0.2465, a localised equilibrium structure (Equilibrium Type III) emerges. In this scenario, Equilibrium Type II becomes unstable and a natural transition to Equilibrium Type III is observed.

If the externally applied shear stress $\tau^0_{\text{ext}}$ is non-negligible, two possible equilibrium patterns are found and the one with larger pair width value corresponds to the stable configuration as suggested by the shaded region in Fig.~\ref{fig_contour_tau_ext}.

In the continuum limit, the two field variables introduced evolve on different time scales. On the faster scale, the dislocation pairs arrange themselves in local structures to satisfy the leading-order force balance. On the slower scale, the pair density evolves driven by the stress gradient, which is a higher-order effect. Consequently, the dipole dynamics, if viewed at the continuum level, can be modelled by an equilibrium equation for $\zeta$ given by Eq.~\eqref{eqn_zeta_static} and an evolution equation for $\phi$ given by Eq.~\eqref{eqn_phi_evo_ts}. All analytical results have been justified through comparison with the underlying DDD simulation results.

\subsection{Implication to the formation of PSBs\label{Sec_discussion_PSBs}}
The finding of a natural transition between equilibrium configurations of dislocations in this paper may shed light on understanding how localised persistent slip band structures emerge within a non-localised channel-vein structure in cyclicly loaded crystals. The analytical results in \S~\ref{Sec_fb_leading_order_zero} suggest that such a transition takes place, when the slip plane spacings drop to a certain value such that the quantity equivalent to $\phi'S$ falls below $0.2465$. In a cyclicly loaded crystal, it is widely recognised that the gaps between slip planes do get narrower as a result of the cross-slip motion of the screw segments in the channels shown in Fig.~\ref{fig_vein_PSBs}(a) (see \cite{Mughrabi_1979, Zhu_MSEA2014}). Nevertheless, the transition in equilibrium patterns due to instability found here may not provide a full explanation to the formation of PSBs, because the PSB walls consist more likely of several dislocation pairs rather than a single pair as indicated by the Equilibrium Type III.

\subsection{Implication to incorporating SSDs into continuum models of plasticity}
The approaches used here to separate physical processes according to their associated time scales also provide us some hints towards incorporating statistically stored dislocations into continuum models of plasticity consistent with the underlying discrete dislocation dynamics. Given $t$ the time scale associated with the continuum model (termed as the continuum time scale), it has been shown that the mutual adjustment within dislocation pairs characterised by the evolution of $\zeta$ takes place so fast that only its steady (equilibrium) state is observable at the continuum time scale. On the other hand, the evolution of the pair density potential $\phi$ takes place so slowly that it appears almost unchanged observed at the continuum time scale. Analogously, a well-established continuum model of plasticity is expected to be hierarchic in time. It should consist of a set of evolution equations for the geometrically necessary dislocations (GNDs) changing at a normal speed accompanied by another set of quasi-static equations describing the SSD structures in equilibrium.

\bibliography{mybib}
\bibliographystyle{siam}

\end{document}